\PassOptionsToPackage{unicode}{hyperref}
\PassOptionsToPackage{hyphens}{url}
\PassOptionsToPackage{dvipsnames,svgnames,x11names}{xcolor}
\documentclass[12pt,a4paper]{article}
\usepackage[top=1.5in, bottom=1.5in, left=1.2in, right=1.2in]{geometry}

\usepackage{amsmath,amssymb}
\usepackage{xr-hyper}
\usepackage{paralist}
\usepackage{iftex}
\newcommand{\proglang}[1]{\textsf{#1}}
\ifPDFTeX
  \usepackage[T1]{fontenc}
  \usepackage[utf8]{inputenc}
  \usepackage{textcomp}
\else 
  \usepackage{unicode-math}
  \defaultfontfeatures{Scale=MatchLowercase}
  \defaultfontfeatures[\rmfamily]{Ligatures=TeX,Scale=1}
\fi
\usepackage{lmodern}
\ifPDFTeX\else  
\fi
\IfFileExists{upquote.sty}{\usepackage{upquote}}{}
\IfFileExists{microtype.sty}{
  \usepackage[]{microtype}
  \UseMicrotypeSet[protrusion]{basicmath} 
}{}
\makeatletter
\@ifundefined{KOMAClassName}{
  \IfFileExists{parskip.sty}{
    \usepackage{parskip}
  }{
    \setlength{\parindent}{0pt}
    \setlength{\parskip}{6pt plus 2pt minus 1pt}}
}{
  \KOMAoptions{parskip=half}}
\makeatother
\usepackage{xcolor}
\usepackage{enumitem}
\usepackage{booktabs, microtype, rotating}
\makeatletter
\ifx\paragraph\undefined\else
  \let\oldparagraph\paragraph
  \renewcommand{\paragraph}{
    \@ifstar
      \xxxParagraphStar
      \xxxParagraphNoStar
  }
  \newcommand{\xxxParagraphStar}[1]{\oldparagraph*{#1}\mbox{}}
  \newcommand{\xxxParagraphNoStar}[1]{\oldparagraph{#1}\mbox{}}
\fi
\ifx\subparagraph\undefined\else
  \let\oldsubparagraph\subparagraph
  \renewcommand{\subparagraph}{
    \@ifstar
      \xxxSubParagraphStar
      \xxxSubParagraphNoStar
  }
  \newcommand{\xxxSubParagraphStar}[1]{\oldsubparagraph*{#1}\mbox{}}
  \newcommand{\xxxSubParagraphNoStar}[1]{\oldsubparagraph{#1}\mbox{}}
\fi
\makeatother

\usepackage{longtable,booktabs,array}
\usepackage{calc}
\usepackage{etoolbox}
\makeatletter
\patchcmd\longtable{\par}{\if@noskipsec\mbox{}\fi\par}{}{}
\makeatother
\IfFileExists{footnotehyper.sty}{\usepackage{footnotehyper}}{\usepackage{footnote}}
\makesavenoteenv{longtable}
\usepackage{graphicx}

\makeatletter
\def\maxwidth{\ifdim\Gin@nat@width>\linewidth\linewidth\else\Gin@nat@width\fi}
\def\maxheight{\ifdim\Gin@nat@height>\textheight\textheight\else\Gin@nat@height\fi}
\makeatother
\setkeys{Gin}{width=\maxwidth,height=\maxheight,keepaspectratio}
\makeatletter
\def\fps@figure{htbp}
\makeatother

\makeatletter
\@ifpackageloaded{caption}{}{\usepackage{caption}}
\AtBeginDocument{%
\ifdefined\contentsname
  \renewcommand*\contentsname{Table of contents}
\else
  \newcommand\contentsname{Table of contents}
\fi
\ifdefined\listfigurename
  \renewcommand*\listfigurename{List of Figures}
\else
  \newcommand\listfigurename{List of Figures}
\fi
\ifdefined\listtablename
  \renewcommand*\listtablename{List of Tables}
\else
  \newcommand\listtablename{List of Tables}
\fi
\ifdefined\figurename
  \renewcommand*\figurename{Figure}
\else
  \newcommand\figurename{Figure}
\fi
\ifdefined\tablename
  \renewcommand*\tablename{Table}
\else
  \newcommand\tablename{Table}
\fi
}
\@ifpackageloaded{float}{}{\usepackage{float}}
\floatstyle{ruled}
\@ifundefined{c@chapter}{\newfloat{codelisting}{h}{lop}}{\newfloat{codelisting}{h}{lop}[chapter]}
\floatname{codelisting}{Listing}

\makeatother
\makeatletter
\@ifpackageloaded{caption}{}{\usepackage{caption}}
\@ifpackageloaded{subcaption}{}{\usepackage{subcaption}}
\makeatother

\ifLuaTeX
  \usepackage{selnolig} 
\fi
\usepackage[]{natbib}
\usepackage{setspace}
\usepackage{dsfont}
\newtheorem{lemma}{Lemma}
\newtheorem{theorem}{Theorem}
\usepackage{bookmark}

\IfFileExists{xurl.sty}{\usepackage{xurl}}{}
\hypersetup{
  pdftitle={Title},
  pdfauthor={Author 1; Author 2},
  pdfkeywords={3 to 6 keywords, that do not appear in the title},
  colorlinks=true,
  linkcolor={blue},
  filecolor={Maroon},
  citecolor={Blue},
  urlcolor={Blue},
  pdfcreator={LaTeX via pandoc}}

\newcommand{\anon}{1}

\graphicspath{{plots}}

\begin{document}

\date{}



\if1\anon
{
\title{\bf White noise testing for functional time series via functional quantile autocorrelation}
\author{Ángel López-Oriona\thanks{Postal address: CEMSE Division, Statistics Program, King Abdullah University of Science and Technology (KAUST), Thuwal 23955-6900, Saudi Arabia. E-mail: angel.lopezoriona@kaust.edu.sa}\hspace{.2cm}\\
Statistics Program \\ 
King Abdullah University of Science and Technology\\
Thuwal, Saudi Arabia \\
\\
Ying Sun \\
Statistics Program \\ 
King Abdullah University of Science and Technology\\
Thuwal, Saudi Arabia \\
\\
Han Lin Shang \\
Department of Actuarial Studies and Business Analytics\\ Macquarie University \\
Sydney, Australia \\}
\maketitle
} \fi

\if0\anon
{
\bigskip
\bigskip
\bigskip
\begin{center}
{\LARGE\bf White noise testing for functional time series via functional quantile autocorrelation}
\end{center}

\medskip
} \fi

\bigskip
\begin{abstract}
We introduce a novel class of nonlinear tests for serial dependence in functional time series, grounded in the functional quantile autocorrelation framework. Unlike traditional approaches based on the classical autocovariance kernel, the functional quantile autocorrelation framework leverages quantile-based excursion sets to robustly capture temporal dependence within infinite-dimensional functional data, accommodating potential outliers and complex nonlinear dependencies. We propose omnibus test statistics and study their asymptotic properties under both known and estimated quantile curves, establishing their asymptotic distribution and consistency under mild assumptions. In particular, no moment conditions are required for the validity of the tests. Extensive simulations and an application to high-frequency financial functional time series demonstrate the methodology's effectiveness, reliably detecting complex serial dependence with a power superior to several existing tests. This work expands the functional time series toolkit, providing a robust framework for inference in settings where traditional methods may fail.

\vspace{.1in}
\vfill
\end{abstract}

\newpage
\doublespacing

\section{Introduction}\label{sectionintroduction}
	
Functional data analysis is an area of statistics that concerns the study of data in graphical representation forms of curves, images, or surfaces that vary over a continuum, such as time, space, or wavelength. Instead of treating each observation as a finite-dimensional vector, this framework regards each observation as a function, enabling inference, modeling, and forecasting that capture the rich, intrinsically infinite-dimensional nature of modern datasets. Applications include the analysis of weather curves, growth trajectories, and spectrometric profiles, where information is often recorded densely and continuously. For detailed introductions and foundational results, refer to \cite{ramsay2005functional}.


In many settings, these functional observations are collected sequentially, giving rise to a functional time series in which the goal is to model and analyze temporal dependence across entire curves observed over time. Important examples include daily pollutant concentration profiles, intraday price curves in finance, and physiological signals in biomedical studies. The development of statistical methodology for functional time series has evolved rapidly in recent years, addressing challenges unique to the interplay between serial dependence and the infinite-dimensional nature of the data; see \cite{HK12} and \cite{KR17} for comprehensive overviews.

A fundamental task in time series analysis is to determine whether observations exhibit serial dependence. This initial diagnostic is crucial for understanding dynamic properties and for guiding model selection and assessment. Analysts commonly evaluate temporal dependence using statistics such as the sample autocorrelation function, supported by graphical tools, and apply hypothesis tests to detect temporal patterns or verify whether residuals resemble uncorrelated noise. When the observations are temporally-dependent curves,  assessing dependence becomes more challenging due to their infinite-dimensional nature. Standard scalar autocorrelation measures may miss important structure within the functions themselves. Specialized methodologies have recently been developed to quantify and test the dependence between functions observed at different time points. In this context, procedures based on relevant functional statistics, such as functional autocorrelation function (FACF) \citep{kokoszka2017inference} or functional spherical autocorrelation function (FSACF) \citep{yeh2023functional}, among others, are essential for rigorous model selection and evaluation and the development of reliable inferences for functional time series.

Despite the growing number of procedures for testing serial dependence in functional time series \citep{kim2023white}, many existing methods are primarily based on second-order or $L^2$-type summaries of the curves, such as covariance operators, (linear) FACF, or projections onto a small number of principal components; see Sections 3 and 4 in \citet{kim2023white}. As a consequence, their performance may deteriorate in the presence of outlying trajectories or heavy-tailed noise, and they are mainly sensitive to dependence encoded in the linear correlations within the mean–covariance structure of the process. Moreover, by focusing on this linear and moment-based structure, these tests can miss more complex forms of serial dependence, for instance, when temporal dependence is concentrated in specific parts of the conditional distribution or in the tails of the functional observations. These limitations motivate the use of quantile-based measures of dependence that are less sensitive to extreme trajectories and can capture distributional and tail behavior beyond first moments. In this spirit, we build on the functional quantile autocorrelation (FQA) introduced in \cite{lopezoriona2025}, where it was proposed as a powerful dependence measure for fuzzy clustering of functional time series. While FQA was previously used in a purely computational setting, the present work develops its asymptotic properties and leverages it for formal hypothesis testing. Hence, we propose FQA as the core ingredient of a rigorous inference framework for assessing serial dependence in functional time series. 

In a nutshell, this work revolves around using FQA to test 
\begin{equation*}
H_0: \{\mathcal{X}_t, t\in\mathbb{Z}\} \text{ is an i.i.d.\ sequence of random elements in } L^2([0,1]),
\end{equation*}
based on a realization of the functional stochastic process $\{\mathcal{X}_t, t\in\mathbb{Z}\}$, against broad alternatives with serial dependence at one or more lags. More specific null hypotheses are given in Section~\ref{sectiontests}. In this context, the main contributions can be summarized as follows:
\begin{asparaitem}
\item We formalize FQA for strictly stationary functional time series in a testing context and investigate its basic statistical properties. In contrast to classical autocorrelation-based methods, which focus on second-order structure, FQA is able to capture more general forms of serial dependence by tracking the temporal behavior of quantile-defined excursion sets.
\item We construct test statistics based on FQA for testing serial dependence in functional time series, with particular emphasis on an omnibus statistic that, at a fixed lag, aggregates FQA over a grid of probability levels and thresholds.
\item We derive the asymptotic distribution of the proposed test statistics under the null hypothesis of serial independence and establish the consistency of the resulting tests under stationarity and ergodicity.
\item We investigate the finite-sample performance of the proposed omnibus test through an extensive simulation study, discuss its advantages over alternative procedures, and illustrate its practical usefulness with an application to financial time series data.
\end{asparaitem}

The remainder of the paper is structured as follows. Section~\ref{sectionrelatedworks} reviews related work on serial dependence testing for functional time series. Section~\ref{sectionbackground} introduces the background and formal definitions of FQA and the associated omnibus test statistic. In Section~\ref{sectionasymptotics}, we study the asymptotic properties of the estimated FQA and of the omnibus statistic. Section~\ref{sectiontests} formally describes the construction of new serial dependence tests based on the proposed methodology. Section~\ref{sectionsimulations} analyzes the finite-sample performance of the omnibus test through an extensive simulation study, while Section~\ref{sectionrealdata} illustrates its application to financial functional time series. Finally, Section~\ref{sectiondiscussion} discusses the implications of the results and outlines potential directions for future research. Proofs of the main theoretical results, together with additional material on the analyses in Section~\ref{sectionsimulations} and the application in Section~\ref{sectionrealdata}, are provided in the Supplement.

\section{Related works}\label{sectionrelatedworks}

White noise testing for functional time series has been the subject of extensive recent research \citep{kim2023white}, which systematically classifies available procedures into several groups. In that framework, tests are often organized according to both the domain in which they are constructed (time domain via, e.g., autocovariance operators, versus spectral domain via, e.g., spectral density operators) and the underlying null hypothesis, distinguishing between independent and identically distributed (i.i.d.)\ white noise (often called strong white noise) and merely uncorrelated functional white noise. This dichotomy is important in practice: testing for strong white noise, the focus of this paper, corresponds to assessing the stricter i.i.d.\ assumption and thereby provides a general check for temporal dependence of both linear and nonlinear types, whereas tests for uncorrelated white noise only target the absence of linear autocorrelation and are therefore particularly suited to residual diagnostics when more complex dependence is expected under the null. A broad collection of these tests, including both time and frequency domain procedures for both types of white noise, is implemented in the \proglang{R} package \textbf{wwntests} \citep{wwntests}, which was developed as a companion to the review of \cite{kim2023white}. Additional white-noise and goodness-of-fit tests for functional time series are provided in the \proglang{R} package \textbf{FTSgof} \citep{ftsgof}, giving practitioners an integrated toolbox for assessing serial dependence and model adequacy in functional time series applications.

Time domain white noise tests for functional time series include procedures targeting both i.i.d.\ white noise and uncorrelated white noise. Tests for i.i.d.\ white noise comprise the Box–Ljung–Pierce-type statistic of \cite{gabrys2007portmanteau}, which is based on the sum of the sample autocorrelation matrices computed from projections of the functional time series, and the fully functional Box–Ljung–Pierce-type statistic of \cite{horvath2013test}, built from \(L^2\)-norms of sample autocovariance functions. In addition, \cite{yeh2023functional} introduces the FSACF, which defines a robust notion of autocorrelation based on measuring the inner product between projections of lagged pairs of the series onto the unit sphere, leading to portmanteau-type statistics that are robust to outliers and lower order moments in the data-generating process. Time domain tests for uncorrelated white noise include the approach of \cite{kokoszka2017inference}, which develops methods for inference on the lagged autocovariance operators of stationary functional time series that are valid under general conditional heteroscedasticity. The framework of \cite{bucher2023portmanteau} proposes a supremum-type statistic involving autocovariance norms to detect departures from white noise in locally stationary settings.

In the spectral domain, tests for i.i.d.\ white noise include the procedure of \cite{characiejus2020general}, which measures the Hilbert--Schmidt distance between a kernel lag-window-type estimator of the spectral density operator of a functional process and the spectral density operator of a white noise process, as well as the characteristic-function-based approach of \cite{hlavka2021testing}, which builds on the key idea that independence is equivalent to the factorization of the joint characteristic function. For uncorrelated white noise, \cite{zhang2016white} develops a Cramér--von Mises–type statistic based on the functional periodogram and derives its null distribution under general uncorrelated but potentially GARCH-like dependence, while \cite{bagchi2018simple} propose a simple test based on an explicit representation of the \(L^2\)-distance between the spectral density operator and its best (\(L^2\))-approximation by a spectral density operator corresponding to a white noise process.

In the more general context of random objects (which include functional data), there are some works worth mentioning in connection with this paper. First, our approach shares certain similarities with the distance profiles introduced by \cite{dubey2024metric}, although the testing framework and statistics proposed here are fundamentally different from those based on distance profiles. Second, a novel method for testing serial independence of object-valued time series in metric spaces is proposed by \cite{jiang2024testing}. This method is built upon the concept of autodistance covariance and employs a Cramér–von Mises-type test statistic that considers a generalized spectral density function. The critical values of the test are approximated using a wild bootstrap procedure.

Our proposal complements the above methods by introducing a time domain test for i.i.d.\ functional white noise that is based on FQA rather than second-order features. By exploiting excursion sets relative to marginal quantile functions and aggregating information over grids of probability levels and thresholds, the resulting omnibus statistic provides a flexible and robust tool for detecting general forms of serial dependence in functional time series. Throughout the manuscript, for simplicity, we assume that each function is observed on a finite grid of points. However, our asymptotic results ensure that the proposed testing framework remains valid in the standard case of infinite-dimensional curves, thereby accounting for the inherent smoothness of functional time series. In summary, our framework applies to both raw and smoothed functional time series, offering an advantage over several alternative approaches.

\section{Functional quantile autocorrelation}\label{sectionbackground}

Following \cite{lopezoriona2025}, we introduce the concepts of functional quantile autocovariance and FQA, and subsequently define an omnibus test statistic based on an estimator of the latter.

Let $\mathcal{I} = [0,1]$, and let $\mathcal{H} = L^2(\mathcal{I})$ denote the Hilbert space of square-integrable functions on $\mathcal{I}$, equipped with the inner product $\langle f, g \rangle_\mathcal{H} = \int_0^1 f(u)\, g(u)\, du$, and the norm $\|f\|_\mathcal{H} = \langle f, f \rangle_\mathcal{H}^{1/2}$. Consider a strictly stationary functional stochastic process \(\{\mathcal{X}_t, t \in \mathbb{Z}\}\), where each \(\mathcal{X}_t\) is a measurable random element in \(\mathcal{H}\); that is, for each fixed \(u \in \mathcal{I}\), \(\mathcal{X}_t(u)\) is a real-valued random variable.

Let $F_{\mathcal{X}(u)}$ denote the marginal cumulative distribution function (CDF) of $\mathcal{X}_t(u)$. The \emph{functional quantile} of level $\tau \in (0,1)$ is defined pointwise as
\begin{equation}\label{fquantile}
q_\tau(u) = \inf \{ x \in \mathbb{R} : F_{\mathcal{X}(u)}(x) \geq \tau \}, \quad u \in \mathcal{I}.
\end{equation}
	
Given $f, g \in \mathcal{H}$, define the excursion set $\mathcal{A} = \{ u \in \mathcal{I} : f(u) \leq g(u) \}$, and let $\mathcal{L}(\mathcal{A})$ denote its Lebesgue measure. Note that since \(f\) and \(g\) admit measurable representatives in \(L^2(\mathcal{I})\), the difference function \(h= f - g\) is measurable. Hence, \(\mathcal{A} = h^{-1}\Big((-\infty, 0]\Big)\) is the pre-image of a Borel-measurable set under a measurable function, implying that \(\mathcal{A}\) is Lebesgue measurable and \(\mathcal{L}(\mathcal{A})\) is well-defined.
	
Given a threshold $\kappa \in (0,1)$, define the binary indicator
\begin{equation*}
\mathds{1}_\kappa(f, g) =
\begin{cases}
1 & \text{if } \mathcal{L}(\mathcal{A}) \leq \kappa, \\
0 & \text{otherwise}.
\end{cases}
\end{equation*}
	
For quantile levels $(\tau, \tau') \in (0,1)^{2}$, thresholds $(\beta, \beta') \in (0,1)^2$, and lag $l \in \mathbb{Z}$, the \emph{functional quantile autocovariance} is given by
\begin{equation*}
\gamma(\tau, \tau', l, \beta, \beta') = \operatorname{Cov}\left [\mathds{1}_\beta(\mathcal{X}_t, q_\tau),\ \mathds{1}_{\beta'}(\mathcal{X}_{t+l}, q_{\tau'}) \right].
\end{equation*}
This quantity, which can be seen as an extension of the classical quantile autocovariance to the functional case, can also be expressed as
\begin{align*}
\gamma(\tau, \tau', l, \beta, \beta') = 
\mathbb{P}\Big(\mathcal{L}(\mathcal{A}_{\tau}^t) \leq \beta,\, \mathcal{L}(\mathcal{A}_{\tau'}^{t+l}) \leq \beta' \Big)-\mathbb{P}\Big(\mathcal{L}(\mathcal{A}_{\tau}^{t}) \leq \beta \Big) \,
\mathbb{P}\Big(\mathcal{L}(\mathcal{A}_{\tau'}^{t+l}) \leq \beta' \Big),
\end{align*}
where $\mathcal{A}_\alpha^{t} = \{ u \in \mathcal{I} : \mathcal{X}_t(u) \leq q_\alpha(u) \}$ for $\alpha \in (0,1)$. Under standard assumptions (e.g., joint measurability of the functional process), the mapping $u \mapsto q_\alpha(u)$ is measurable for each fixed quantile level $\alpha \in (0,1)$. This ensures that the excursion set $\mathcal{A}_\alpha^{t}$ is Lebesgue measurable, and its measure $\mathcal{L}(\mathcal{A}_\alpha^t)$ is well-defined. Note that $\mathcal{L}(\mathcal{A}_\tau^t)$ provides a natural way of quantifying the proportion of points in $\mathcal{I}$ for which the curve $\mathcal{X}_t$ lies below $q_\tau$. Hence, the joint distribution of $\mathcal{L}(\mathcal{A}_\tau^t)$ and $\mathcal{L}(\mathcal{A}_{\tau'}^{t+l})$ provides a meaningful description of the serial dependence structure of $\{\mathcal{X}_t, t \in \mathbb{Z}\}$ for given quantile curves $q_\tau$ and $q_{\tau'}$. In fact, for fixed $\tau$ and $\tau'$, $\gamma(\tau, \tau', l, \beta, \beta')$ can be seen, as a function of $(\beta, \beta') \in (0,1)^2$, as the difference between this joint distribution and the joint distribution of $\mathcal{L}(\mathcal{A}_\tau^t)$ and $\mathcal{L}(\mathcal{A}_{\tau'}^{t+l})$ in the case of independence. Thus,a rich picture of the serial dependence patterns exhibited by $\{\mathcal{X}_t, t \in \mathbb{Z}\}$ can be obtained by examining $\gamma(\tau, \tau', l, \beta, \beta')$ for multiple values of $\tau, \tau'$ and $\beta, \beta'$.
	
The FQA is defined as
\begin{small}
\begin{equation*}
\rho(\tau, \tau', l, \beta, \beta') =  \frac{\gamma(\tau, \tau', l, \beta, \beta')}{\bigg[\mathbb{P}\Big(\mathcal{L}(\mathcal{A}_\tau^t)\leq \beta\Big)\,\Big(1-\mathbb{P}\big(\mathcal{L}(\mathcal{A}_\tau^t)\leq \beta\big)\Big)\,\mathbb{P}\Big(\mathcal{L}(\mathcal{A}_{\tau'}^{t+l})\leq \beta'\Big)\,\Big(1-\mathbb{P}\big(\mathcal{L}(\mathcal{A}_{\tau'}^{t+l})\leq \beta'\big)\Big) \bigg]^{1/2}}.
\end{equation*}
\end{small}
	
From the above definition, it is clear that $-1 \leq \rho(\tau, \tau', l, \beta, \beta') \leq 1$ and $\rho(\tau, \tau', l, \beta, \beta')=\rho(\tau', \tau, -l, \beta', \beta)$. In addition, we have that FQA is well-defined for any strictly stationary process in $\mathcal{H}$, and no moment conditions are required for its validity. 
	
In practice, the previous quantities must be estimated from $T$-length realizations of the functional stochastic process $\{\mathcal{X}_t,\, t \in \mathbb{Z}\}$, $\boldsymbol{\mathcal{X}}=(\mathcal{X}_1, \mathcal{X}_2, \ldots, \mathcal{X}_T)$, often referred to as \textit{functional time series}. We consider that each function in the realization is observed in a common set of $p$ evenly spaced points, $\mathfrak{U} = \{u_1, u_2, \ldots, u_p\} \subset [0, 1]$, with $u_1=0$ and $u_p=1$. Therefore, the information in the realization $\boldsymbol{\mathcal{X}}$ can be expressed through the matrix:
\begin{equation*}
\boldsymbol{\mathcal{X}}_\text{M} = 
\begin{pmatrix}
\mathcal{X}_1(u_1) & \mathcal{X}_1(u_2) & \cdots & \mathcal{X}_1(u_p) \\
\mathcal{X}_2(u_1) & \mathcal{X}_2(u_2) & \cdots & \mathcal{X}_2(u_p) \\
\vdots   & \vdots   & \ddots & \vdots   \\
\mathcal{X}_T(u_1) & \mathcal{X}_T(u_2) & \cdots & \mathcal{X}_T(u_p)
\end{pmatrix},
\end{equation*}
where $\mathcal{X}_t(u_j)$ is the value of the function $\mathcal{X}_t$ at the point $u_j$, $t=1,\ldots,T$, $j=1,\ldots,p$. Note that estimators of $\gamma(\tau, \tau', l, \beta, \beta')$ and $\rho(\tau,\tau',l,\beta,\beta')$ can be computed by estimating the probabilities of $\mathbb{P}\Big(\mathcal{L}(\mathcal{A}_\tau^t) \leq \beta\Big)$ and $\mathbb{P}\Big(\mathcal{L}(\mathcal{A}_\tau^t) \leq \beta,\, \mathcal{L}(\mathcal{A}_{\tau'}^{t+l}) \leq \beta'\Big)$ in a natural way as
\begin{align*}
\widehat{\mathbb{P}}\Big(\mathcal{L}(\mathcal{A}_\tau^t) \le \beta\Big) &= \frac{1}{T}\sum_{i=1}^{T}\mathbb{I}\bigg(\frac{\# \widehat{\mathcal{A}}_{\tau}^{i}}{p} \le \beta\bigg),\\
\widehat{\mathbb{P}}\Big(\mathcal{L}(\mathcal{A}_\tau^t)\le \beta,\mathcal{L}(\mathcal{A}^{t+l}_{\tau'})\le \beta' \Big) &=\frac{1}{T}\sum_{i=1}^{T-l}\mathbb{I}\bigg(\frac{\# \widehat{\mathcal{A}}_{\tau}^{i}}{p} \le \beta\bigg)\mathbb{I}\bigg(\frac{\# \widehat{\mathcal{A}}_{\tau'}^{i+l}}{p} \le \beta'\bigg),
\end{align*}
respectively, where $\mathbb{I}(\cdot)$ denotes the classical binary indicator function, the notation $\#$ is used to indicate the cardinal of a set, and $\widehat{\mathcal{A}}_{\alpha}^{k}=\{u \in \mathfrak{U}: \mathcal{X}_k(u) \le \widehat{q}_\alpha(u)\}$ for $k=1, \ldots, T$ and $\alpha \in (0,1)$, with $\widehat{q}_\alpha(\cdot)$ being a standard estimator of the $\alpha$\textsuperscript{th} quantile of a real functional random variable as defined in~\eqref{fquantile}. The corresponding estimators for $\gamma(\tau, \tau', l, \beta, \beta')$ and $\rho(\tau, \tau', l, \beta, \beta')$ are denoted as $\widehat{\gamma}(\tau, \tau', l, \beta, \beta')$ and $\widehat{\rho}(\tau, \tau', l, \beta, \beta')$, respectively. Throughout the paper, we assume that each function is observed in a sufficiently dense grid of points $\mathfrak{U}$ so that discretization errors can be neglected.
	
The estimator $\widehat{\rho}(\tau, \tau', l, \beta, \beta')$ can be used as a test statistic for detecting serial dependence in a functional time series at a given lag $l \in \mathbb{Z}$, for fixed probability levels $(\tau, \tau')$ and thresholds $(\beta, \beta')$. To aggregate information on dependence across a range of parameters, we consider the omnibus statistic
\begin{equation}\label{is}
W_T(l)
= \int_0^1 \int_0^1 \int_0^1 \int_0^1
\widehat{\rho}^2(\tau, \tau', l, \beta, \beta') \, d\tau \, d\tau' \, d\beta \, d\beta'.
\end{equation}

Because $W_T(l)$ is built directly from the estimated FQA, which in turn is defined through quantile curves and excursion sets, a test based on this statistic is expected to mitigate several drawbacks of most existing white noise tests for functional time series (see Sections~\ref{sectionintroduction} and \ref{sectionrelatedworks}). In particular, it should exhibit greater robustness to heavy tails and outliers, and an enhanced ability to capture complex forms of functional dependence. Moreover, since FQA does not rely on moment assumptions, no finite-moment conditions are required for testing via $W_T(l)$, so the resulting testing framework remains valid under relatively minimal conditions beyond strict stationarity (refer to Section~\ref{sectionasymptotics}).

\section{Asymptotic properties of the sample FQA and the omnibus statistic}\label{sectionasymptotics}
	
This section is devoted to examining the asymptotic behavior of the sample FQA and the integral-type statistic.
	
\subsection{Asymptotic behavior of the sample FQA and the omnibus statistic under independence when the quantile curve is known}\label{sectionad1}
	
We now provide the asymptotic behavior of the FQA estimator in Section~\ref{sectionbackground}, under the null hypothesis that \(\{\mathcal{X}_t, t \in \mathbb{Z}\}\) is a sequence of i.i.d. random functions. Although in applied settings, the quantile function curve \( q_\tau \) must inevitably be estimated from observed data to construct the sample FQA, it is conceptually helpful to begin by examining the properties of FQA estimators in the hypothetical scenario where \( q_\tau \) is fully known. This strategy allows us to isolate and understand which attributes of the statistical estimators and the stochastic process \(\{\mathcal{X}_t, t \in \mathbb{Z}\}\) are fundamental to inference about FQA itself, without conflating them with the additional uncertainty introduced by the estimation of quantile curves. In the subsequent section, we explore the statistical properties when \( q_\tau \) is replaced by a consistent sample-based estimator. 
	
Assuming $q_\tau$ is known for all $\tau \in (0,1)$, a natural estimator of FQA is given by
\begin{small}
\begin{equation}\label{fqaqknown}
\widetilde{\rho}(\tau, \tau', l, \beta, \beta') =  \frac{\widetilde{\mathbb{P}}\Big(\mathcal{L}(\mathcal{A}_{\tau}^t) \leq \beta,\, \mathcal{L}(\mathcal{A}_{\tau'}^{t+l}) \leq \beta' \Big)-\widetilde{\mathbb{P}}\Big(\mathcal{L}(\mathcal{A}_{\tau}^{t}) \leq \beta \Big) \,
\widetilde{\mathbb{P}}\Big(\mathcal{L}(\mathcal{A}_{\tau'}^{t+l}) \leq \beta' \Big)}{\bigg[\widetilde{\mathbb{P}}\Big(\mathcal{L}(\mathcal{A}_\tau^t)\leq \beta\Big)\,\Big(1-\widetilde{\mathbb{P}}\big(\mathcal{L}(\mathcal{A}_\tau^t)\leq \beta\big)\Big)\,\widetilde{\mathbb{P}}\Big(\mathcal{L}(\mathcal{A}_{\tau'}^{t+l})\leq \beta'\Big)\,\Big(1-\widetilde{\mathbb{P}}\big(\mathcal{L}(\mathcal{A}_{\tau'}^{t+l})\leq \beta'\big)\Big) \bigg]^{1/2}},
\end{equation}
\end{small}
with 
\begin{align*}
\widetilde{\mathbb{P}}\Big(\mathcal{L}(\mathcal{A}_\tau^t) \le \beta\Big) &= \frac{1}{T}\sum_{i=1}^{T}\mathbb{I}\bigg(\frac{\# \widetilde{\mathcal{A}}_{\tau}^{i}}{p} \le \beta\bigg),\\
\widetilde{\mathbb{P}}\Big(\mathcal{L}(\mathcal{A}_\tau^t)\le \beta,\mathcal{L}(\mathcal{A}^{t+l}_{\tau'})\le \beta' \Big) &=\frac{1}{T}\sum_{i=1}^{T-l}\mathbb{I}\bigg(\frac{\# \widetilde{\mathcal{A}}_{\tau}^{i}}{p} \le \beta\bigg)\mathbb{I}\bigg(\frac{\# \widetilde{\mathcal{A}}_{\tau'}^{i+l}}{p} \le \beta'\bigg),
\end{align*}
where $\widetilde{\mathcal{A}}_{\alpha}^{k}=\{u \in \mathfrak{U}: \mathcal{X}_k(u) \le q_\alpha(u)\}$ for $k=1, \ldots, T$.
	
Before going into the asymptotics of $\widetilde{\rho}(\tau, \tau', l, \beta, \beta')$, the following lemma shows that the proportion of points in the discrete excursion set converges to the Lebesgue measure of the corresponding continuous excursion set as the observation grid becomes increasingly dense.
\begin{lemma}\label{lemma1}
Let $\boldsymbol{\mathcal{X}}=(\mathcal{X}_1, \mathcal{X}_2, \ldots, \mathcal{X}_T)$ be a realization of the stochastic process \(\{\mathcal{X}_t, t \in \mathbb{Z}\}\) and fix $\tau \in (0,1)$. For $k=1,\ldots, T$, we have
\begin{equation*}
\lim_{p \to \infty}\frac{\# \widehat{\mathcal{A}}_\tau^k}{p} = \mathcal{L}\left(\overline{\mathcal{A}}_{\tau}^{k}\right),\qquad
\lim_{p \to \infty}\frac{\# \widetilde{\mathcal{A}}_\tau^k}{p} = \mathcal{L}\left(\breve{\mathcal{A}}_{\tau}^{k}\right),
\end{equation*}
where $\overline{\mathcal{A}}_{\tau}^{k}=\{u \in \mathcal{I}: \mathcal{X}_k(u) \le \widehat{q}_\tau(u)\}$ and $\breve{\mathcal{A}}_{\tau}^{k}=\{u \in \mathcal{I}: \mathcal{X}_k(u) \le q_\tau(u)\}$.
\end{lemma}
	
In Lemma~\ref{lemma1}, standard convergence for a sequence of numbers is considered, since the grid $\mathfrak{U}$ is not random.
	
Let us now consider the following assumptions: 
\begin{enumerate}[label=\textbf{(A\arabic*)}]
\item\label{aiid} The realization $\boldsymbol{\mathcal{X}}=(\mathcal{X}_1, \mathcal{X}_2, \ldots, \mathcal{X}_T)$ consists of i.i.d. random elements in $L^2(\mathcal{I})$, generated from the stochastic process \(\{\mathcal{X}_t, t \in \mathbb{Z}\}\).
\item\label{acdf} For each $u \in \mathcal{I}$, the marginal CDF $F_{\mathcal{X}(u)}$ is continuous and strictly increasing.
\end{enumerate}
	
Assumption~\ref{acdf} ensures that the pointwise quantile function $q_\tau$ is uniquely defined for all $\tau \in (0,1)$. Continuity and strict monotonicity imply that there are no flat regions or discontinuities in the distribution of $\mathcal{X}_t(u)$, so the infimum in the quantile definition always yields a single, well-defined value. As a result, the excursion set $\{ u \in \mathcal{I} : \mathcal{X}_t(u) \leq q_\tau(u) \}$ is unambiguous and its Lebesgue measure is properly defined for all relevant $(t, \tau)$. Furthermore, this assumption facilitates the continuity (with respect to $\tau$) and measurability (with respect to $u$) of the quantile curve, which are necessary for a rigorous probabilistic and statistical analysis of the proposed methodology. In practice, this requirement is generally satisfied when the underlying functional data arise from processes with continuous distributions, a setting commonly encountered in functional data analysis applications.
	
Now, we present the asymptotic distribution of $\widetilde{\rho}(\tau, \tau', l, \beta, \beta')$ under Assumptions~\ref{aiid} and~\ref{acdf}. 
\begin{theorem}\label{th1}
Let $\boldsymbol{\mathcal{X}}=(\mathcal{X}_1, \mathcal{X}_2, \ldots, \mathcal{X}_T)$ be a realization of the stochastic process \(\{\mathcal{X}_t, t \in \mathbb{Z}\}\) and fix $(\tau, \tau') \in (0,1)^2$, $(\beta, \beta') \in (0,1)^2$, and lag $l \geq 1$. Suppose that Assumptions~\ref{aiid} and~\ref{acdf} hold. Then, as $T \to \infty$,
\begin{equation*}
\sqrt{T}\widetilde{\rho}(\tau, \tau', l, \beta, \beta')\xrightarrow{d}\mathcal{N}\left(0,\, \sigma^2 \right),
\end{equation*}
where $\sigma^2$ is a finite asymptotic variance depending on the joint distribution of the underlying indicator variables.
\end{theorem}
	
The notation \(\xrightarrow{d}\) above denotes convergence in distribution. The explicit expression for the asymptotic variance $\sigma^2$ is provided in the proof, but is omitted here for brevity. 
	
To analyze the asymptotic behavior of the integral statistic $W_T(l)$ introduced in~\eqref{is}, for simplicity, we will work with the corresponding (nonnormalized) discretized version of it, again assuming that the quantile curve $q_\tau$ is fully known. Thus, we consider a set of $P$ probability levels, $\mathcal{T} = \{\tau_1, \tau_2, \ldots, \tau_P\}\subset (0,1)$, the set of $B$ thresholds, $\mathcal{B} = \{\beta_1, \beta_2, \ldots, \beta_B\} \subset (0,1)$, and, for a fixed lag $l$, define the statistic
\begin{equation*}
\widetilde{S}_T(l) = \sum_{i,j=1}^{P} \sum_{k,\ell=1}^{B}\widetilde{\rho}^2(\tau_i, \tau_j, l, \beta_k, \beta_\ell).
\end{equation*}
	
The following theorem provides the asymptotic distribution of the statistic $\widetilde{S}_T(l)$ under Assumptions~\ref{aiid} and~\ref{acdf}. 
\begin{theorem}\label{th2}
Let $\boldsymbol{\mathcal{X}}=(\mathcal{X}_1, \mathcal{X}_2, \ldots, \mathcal{X}_T)$ be a realization of the stochastic process \(\{\mathcal{X}_t, t \in \mathbb{Z}\}\) and fix the sets $\mathcal{T} = \{\tau_1, \tau_2, \ldots, \tau_P\}\subset (0,1)$ and $\mathcal{B} = \{\beta_1, \beta_2, \ldots, \beta_B\} \subset (0,1)$, and lag $l \geq 1$. Given $\tau, \tau' \in \mathcal{T}$ and $\beta, \beta' \in \mathcal{B}$, define $\widetilde{\rho}(\tau, \tau', l, \beta, \beta')$ as in~\eqref{fqaqknown}. Suppose that Assumptions~\ref{aiid} and~\ref{acdf} hold. Then, as $T \to \infty$,
\begin{equation*}
T\widetilde{S}_T(l) \xrightarrow{\;\; d \;\;}\sum_{j=1}^{P^2B^2}\lambda_jY_j^2,
\end{equation*}
where $\{Y_j\}_{j=1}^{P^2B^2}$ are independent standard normal random variables and $\{\lambda_j\}_{j=1}^{P^2B^2}$ are nonnegative real numbers. 
\end{theorem}

The explicit expression for $\{\lambda_j\}_{j=1}^{P^2B^2}$ is provided in the proof, but omitted here for brevity. 

\subsection{Asymptotic behavior of the sample FQA and the omnibus statistic under independence with unknown quantile curve}\label{sectionad2}
	
In practical applications, the quantile curve $q_{\tau}$ is not directly observable and must be estimated from a functional time series. A commonly used approach is to construct $\widehat{q}_{\tau}$ as the generalized inverse of the empirical cumulative distribution function (ECDF) constructed from the sample. More precisely, for each \( u \in \mathcal{I} \), we consider
\begin{equation*}
\widehat{q}_\tau(u) = \inf \left\{ x \in \mathbb{R} : 	\widehat{F}_{\mathcal{X}(u)}(x) \geq \tau \right\},
\end{equation*}
where the ECDF at point $u$ is given by
\begin{equation*}
\widehat{F}_{\mathcal{X}(u)}(x) = \frac{1}{T} \sum_{i=1}^T \mathbb{I}\bigl(\mathcal{X}_i(u) \leq x \bigr),
\end{equation*}
with $\{ \mathcal{X}_i(u) \}_{i=1}^{T}$ being the sample paths of the functional observations evaluated at $u$. Based on this estimator for the quantile curve, the sample FQA is constructed following the procedure introduced in Section~\ref{sectionbackground}.
	
Our objective now is to investigate whether the asymptotic properties of 
\(\widehat{\rho}(\tau, \tau', l, \beta, \beta')\) remain valid when the true quantile curves are replaced by their sample estimators. Specifically, we aim to determine sufficient conditions under which the estimation step has an asymptotically negligible effect. To this end, we begin by stating a version of Theorem~\ref{th1} that accounts for quantile estimation, under the following additional regularity assumptions:
	
\begin{enumerate}[label=\textbf{(A\arabic*)},resume]
\item\label{aquantileconv} The estimated quantile curves \(\widehat{q}_{\tau} : \mathcal{I} \to \mathbb{R}\), constructed as indicated above, satisfy
\begin{equation*}
\sup_{\tau \in (0,1)} \sup_{u \in \mathcal{I}} \left| \widehat{q}_{\tau}(u) - q_{\tau}(u) \right| \xrightarrow{p} 0.
\end{equation*}
\end{enumerate}

This assumption is a standard uniform consistency requirement for empirical quantile curves and is satisfied under mild regularity conditions. In particular, when the functional time series consists of i.i.d.\ random functions, Assumption~\ref{acdf} holds, and the sample paths are almost surely continuous as functions of \(u\), functional Glivenko--Cantelli results \citep{van1998asymptotic} imply that the ECDF $\widehat{F}_{\mathcal{X}(u)}(x)$ converge uniformly to $F_{\mathcal{X}(u)}(x)$ over $(u,x) \in \mathcal{I} \times \mathbb{R}$. Since the quantile curves are defined as generalized inverses of these ECDF, the uniform convergence of the ECDF entails the uniform convergence of the corresponding empirical quantile curves, yielding Assumption~\ref{aquantileconv}. This condition ensures that the estimated quantile curves are uniformly consistent. However, this condition alone is not sufficient to guarantee that the effect of quantile estimation is asymptotically negligible. Additional conditions are therefore required to control the impact of the plug-in step. Moreover, the continuity conditions required above are satisfied in many classical models for functional time series. For dependent functional time series, Assumption~\ref{aquantileconv} can be justified under suitable weak dependence conditions, such as mixing-type assumptions, together with mild regularity conditions that are known to yield uniform consistency of the estimated quantile curves.

\begin{enumerate}[label=\textbf{(A\arabic*)},resume]
\item\label{aplugin} For each fixed $(\tau,\tau') \in (0,1)^2$ and $(\beta,\beta') \in (0,1)^2$, the following conditions hold as $T \to \infty$:
\begin{equation*}
\frac{1}{T} \sum_{t=1}^{T}
\left|
\mathbb{I}\Big(\mathcal{L}(\overline{\mathcal{A}}^t_{\tau}) \le \beta\Big)
-
\mathbb{I}\Big(\mathcal{L}(\breve{\mathcal{A}}^t_{\tau}) \le \beta\Big)
\right|
= o_p(T^{-1/2}),
\end{equation*}
and
\begin{equation*}
\frac{1}{T} \sum_{t=1}^{T-l}
\left|
\mathbb{I}\Big(\mathcal{L}(\overline{\mathcal{A}}^t_{\tau}) \le \beta\Big)
\mathbb{I}\Big(\mathcal{L}(\overline{\mathcal{A}}^{t+l}_{\tau'}) \le \beta'\Big)
-
\mathbb{I}\Big(\mathcal{L}(\breve{\mathcal{A}}^t_{\tau}) \le \beta\Big)
\mathbb{I}\Big(\mathcal{L}(\breve{\mathcal{A}}^{t+l}_{\tau'}) \le \beta'\Big)
\right|
= o_p(T^{-1/2}).
\end{equation*}
\end{enumerate}

Assumption~\ref{aplugin} is a high-level stability condition that ensures that the effect of replacing the true quantile curves by their estimators is asymptotically negligible on the scale $T^{-1/2}$. Intuitively, this condition requires that the proportion of time points for which the thresholded excursion indicators differ when using estimated versus true quantile curves is sufficiently small. While this condition is not implied by the uniform consistency of the quantile curves alone, it is plausible in settings where the distribution of the excursion measure $\mathcal{L}(\mathcal{A}^t_{\tau})$ places a limited probability mass in shrinking neighborhoods of the threshold $\beta$, so that small perturbations in the quantile curves are unlikely to alter the corresponding indicator values. In particular, its validity depends on the local behavior of the density of $\mathcal{L}(\mathcal{A}^t_{\tau})$ around $\beta$. This situation is more likely to arise in functional time series where trajectories exhibit strong dependence across the domain, leading to excursion measures $\mathcal{L}(\mathcal{A}^t_{\tau})$ that are sufficiently variable and do not concentrate excessively around the threshold $\beta$. 

Since excursion measures are obtained by integrating indicator functions over the domain, the effect of quantile estimation is averaged across $u \in \mathcal{I}$, which may further stabilize the resulting quantities. Although this aggregation does not eliminate the impact of quantile estimation in general, it suggests that its influence on the thresholded indicators may be limited in many practical situations. A rigorous assessment of Assumption~\ref{aplugin} under general conditions would require a detailed analysis of the joint behavior of the estimated quantile process and the induced excursion measures, particularly in regions where $\mathcal{L}(\mathcal{A}^t_{\tau})$ is close to $\beta$. Such an analysis is technically involved and is beyond the scope of this paper. In practice, for the omnibus statistic, it may be sufficient that this condition holds approximately for most of the quantile levels and thresholds considered. In Section~\ref{sectionsimulations}, we provide empirical evidence supporting the plausibility of this condition in a variety of settings, where the finite-sample behavior of the proposed omnibus test under the null hypothesis is unaffected by the use of estimated quantile curves.

\begin{theorem}\label{th3}
Let $\boldsymbol{\mathcal{X}}=(\mathcal{X}_1, \mathcal{X}_2, \ldots, \mathcal{X}_T)$ be a realization of the stochastic process \(\{\mathcal{X}_t, t \in \mathbb{Z}\}\) and fix $(\tau, \tau') \in (0,1)^2$, $(\beta, \beta') \in (0,1)^2$, and lag $l \geq 1$. Suppose that Assumptions~\ref{aiid}-\ref{aplugin} hold, then \(\widehat{\rho}(\tau, \tau', l, \beta, \beta')\) satisfies Theorem \ref{th1} in place of \(\widetilde{\rho}(\tau, \tau', l, \beta, \beta')\). 
\end{theorem}

Theorem~\ref{th3} shows that, under Assumptions~\ref{aiid}-\ref{aplugin}, the same inference procedures (e.g., confidence regions and hypothesis tests) based on \(\widetilde{\rho}(\tau, \tau', l, \beta, \beta')\) remain asymptotically valid when using \(\widehat{\rho}(\tau, \tau', l, \beta, \beta')\) instead.

A similar result holds for the corresponding version of the statistic $\widetilde{S}_T(l)$ when the quantile curves are unknown and must be estimated. Specifically, given the set of $P$ probability levels, $\mathcal{T} = \{\tau_1, \tau_2, \ldots, \tau_P\}\subset (0,1)$, the set of $B$ thresholds, $\mathcal{B} = \{\beta_1, \beta_2, \ldots, \beta_B\} \subset (0,1)$, and, for a fixed lag $l$, this statistic is defined as
\begin{equation}\label{is_2}
\widehat{S}_T(l)= \sum_{i,j=1}^{P} \sum_{k,\ell=1}^{B}\widehat{\rho}^2(\tau_i, \tau_j, l, \beta_k, \beta_\ell).
\end{equation}

\begin{theorem}\label{th4}
Let $\boldsymbol{\mathcal{X}}=(\mathcal{X}_1, \mathcal{X}_2, \ldots, \mathcal{X}_T)$ be a realization of the stochastic process \(\{\mathcal{X}_t, t \in \mathbb{Z}\}\) and fix the sets $\mathcal{T} = \{\tau_1, \tau_2, \ldots, \tau_P\}\subset (0,1)$ and $\mathcal{B} = \{\beta_1, \beta_2, \ldots, \beta_B\} \subset (0,1)$, and lag $l \geq 1$. Given $\tau, \tau' \in \mathcal{T}$ and $\beta, \beta' \in \mathcal{B}$, define $\widehat{\rho}(\tau, \tau', l, \beta, \beta')$ as indicated in Section~\ref{sectionbackground}. Suppose that Assumptions~\ref{aiid}-\ref{aplugin} hold, then $\widehat{S}_T(l)$ satisfies Theorem~\ref{th2} in place of $\widetilde{S}_T(l)$. 
\end{theorem}
	
\subsection{Consistency of the sample FQA and the omnibus statistic}\label{sectionad3}

We now establish the asymptotic consistency of the sample FQA $\widehat{\rho}(\tau,\tau',l,\beta,\beta')$ and the omnibus statistic $\widehat{S}_T(l)$. For this purpose, we introduce the following assumption:
\begin{enumerate}[label=\textbf{(A\arabic*)},resume]
\item\label{asse} The realization $\boldsymbol{\mathcal{X}}=(\mathcal{X}_1,\ldots,\mathcal{X}_T)$ is generated by a strictly stationary and ergodic stochastic process $\{\mathcal{X}_t, t\in\mathbb{Z}\}$.
\end{enumerate}

\begin{theorem}\label{th5}
Let $\boldsymbol{\mathcal{X}}=(\mathcal{X}_1,\ldots,\mathcal{X}_T)$ 
be a realization of $\{\mathcal{X}_t, t\in\mathbb{Z}\}$, and fix
$(\tau,\tau')\in(0,1)^2$, $(\beta,\beta')\in(0,1)^2$, and lag $l\geq 1$. 
Suppose that Assumptions~\ref{acdf}-\ref{asse} hold. Then $\widehat{\rho}(\tau,\tau',l,\beta,\beta')\xrightarrow{p}
\rho(\tau,\tau',l,\beta,\beta')$ as $T\to\infty$. Moreover, for 
$\mathcal{T}=\{\tau_1,\ldots,\tau_P\}\subset(0,1)$ and 
$\mathcal{B}=\{\beta_1,\ldots,\beta_B\}\subset(0,1)$, the omnibus statistic $\widehat{S}_T(l)$ defined in~\eqref{is_2} satisfies 
\[
\widehat{S}_T(l)\xrightarrow{p} 
S_T(l)=\sum_{i,j=1}^P\sum_{k,\ell=1}^B 
\rho^2(\tau_i,\tau_j,l,\beta_k,\beta_\ell).
\]
\end{theorem}

\section{Construction of serial dependence tests}\label{sectiontests}

We employ the asymptotic results derived in Section~\ref{sectionasymptotics} to develop FQA-based tests for serial dependence in functional time series. In practice (see Sections~\ref{sectionsimulations} and~\ref{sectionrealdata}), our aim is to assess the null hypothesis of strong white noise stated in Assumption~\ref{aiid}, namely \(H_0: \{\mathcal{X}_t, t\in\mathbb{Z}\} \text{ is an i.i.d.\ sequence of random elements in } L^2(\mathcal{I})\), against broad alternatives under which the process displays serial dependence at one or more lags. In this framework, we clarify the specific forms of the underlying null hypotheses when using the sample FQA and the omnibus statistic for testing.

\subsection{Tests at fixed levels and thresholds}

Fix $(\tau,\tau')\in(0,1)^2$, $(\beta,\beta')\in(0,1)^2$, and a positive integer lag $l\geq 1$. We first consider tests for the null hypothesis $H_{0,l}:\ \rho(\tau,\tau',l,\beta,\beta') = 0$, based on the sample FQA $\widehat{\rho}(\tau,\tau',l,\beta,\beta')$ defined in Section~\ref{sectionbackground}. If Assumptions~\ref{aiid}-\ref{aplugin} hold, Theorem~\ref{th3} implies that, under $H_{0,l}$,
\begin{equation*}
\sqrt{T}\,\widehat{\rho}(\tau,\tau',l,\beta,\beta') \xrightarrow{d} \mathcal{N}\bigl(0,\sigma^2(\tau,\tau',l,\beta,\beta')\bigr),
\end{equation*}
where the finite asymptotic variance $\sigma^2(\tau,\tau',l,\beta,\beta')$ is given in the proof of Theorem~\ref{th1}. Consequently, an asymptotically level-$\alpha$ test of $H_{0,l}$ rejects the null for large values of the standardized statistic $\frac{\sqrt{T}\,\widehat{\rho}(\tau,\tau',l,\beta,\beta')}{\widehat{\sigma}(\tau,\tau',l,\beta,\beta')}$, where $\widehat{\sigma}(\tau,\tau',l,\beta,\beta')$ is a consistent estimator of $\sigma(\tau,\tau',l,\beta,\beta')$. In particular, rejecting $H_{0,l}$ whenever $\left|
\frac{\sqrt{T}\,\widehat{\rho}(\tau,\tau',l,\beta,\beta')}
     {\widehat{\sigma}(\tau,\tau',l,\beta,\beta')}
\right|
\;>\;
z_{1-\alpha/2}$, with $z_{1-\alpha/2}$ denoting the $(1-\alpha/2)$ quantile of the standard normal distribution, yields an asymptotically sized $\alpha$ test. For a level of significance~$\alpha$, an associated $(1-\alpha)$ confidence interval for $\rho(\tau,\tau',l,\beta,\beta')$ is 
\[
\widehat{\rho}(\tau,\tau',l,\beta,\beta')
\ \pm\
\frac{\widehat{\sigma}(\tau,\tau',l,\beta,\beta')}{\sqrt{T}}\,z_{1-\alpha/2}.
\]

The foregoing construction focuses on testing serial dependence in a single lag $l$ and for a fixed combination $(\tau,\tau',\beta,\beta')$. In analogy to classical time series analysis, it is also possible to build portmanteau-type statistics by aggregating quantities based on $\widehat{\rho}(\tau,\tau',l,\beta,\beta')$ over a collection of lags. For a fixed positive integer $L$, this corresponds to testing the null hypothesis $H_{0,L}':\ \rho(\tau,\tau',l,\beta,\beta') = 0$ for all $l \in \{1,\ldots,L\}$, and constructing a statistic that aggregates the suitable functions of the FQA estimators across $l$. The asymptotic null behavior of such portmanteau statistics can be studied by extending the arguments developed for the fixed-lag case. These extensions fit naturally within our framework but are not pursued here, as our primary focus is on fixed-lag inference for simplicity.

\subsection{Omnibus tests over grids of parameters}

While tests at a fixed combination $(\tau, \tau',\beta, \beta')$ are useful for targeted inference, in practice, it is desirable to aggregate information over a range of quantile levels and thresholds. To this end, recall the omnibus statistic $\widehat{S}_T(l) = \sum_{i,j=1}^P\sum_{k,\ell=1}^B
\widehat{\rho}^2(\tau_i,\tau_j,l,\beta_k,\beta_\ell)$, constructed from probability levels $\mathcal{T}=\{\tau_1,\ldots,\tau_P\}\subset(0,1)$ and thresholds $\mathcal{B}=\{\beta_1,\ldots,\beta_B\}\subset(0,1)$; see~\eqref{is_2}. Under $H_0$ (Assumption~\ref{aiid}) and Assumptions~\ref{acdf}-\ref{aplugin},
Theorem~\ref{th4} implies that
\begin{equation*}
T\,\widehat{S}_T(l)\ \xrightarrow{d}\ \sum_{j=1}^{P^2B^2}\lambda_j Y_j^2,
\end{equation*}
where $\{Y_j\}_{j=1}^{P^2B^2}$ are independent standard normal random variables, $\{\lambda_j\}_{j=1}^{P^2B^2}$ are nonnegative real numbers determined by the joint distribution of the indicator functions.

The limiting null distribution is thus a finite linear combination of independent chi-squared variables. In practice, the values $\lambda_j$ are unknown and must be estimated from the data. Let us denote $\widehat{\boldsymbol{\rho}}_T(l)
=
\Bigl(
\widehat{\rho}(\tau_i,\tau_j,l,\beta_k,\beta_\ell)
\Bigr)_{1\leq i,j\leq P,\ 1\leq k,\ell\leq B}
\in\mathbb{R}^{P^2B^2}$, as the stacked vector of FQA estimators over the chosen grids. Under $H_0$, arguments in the proofs of Theorems~\ref{th2} and \ref{th4} show that
\begin{equation*}
\sqrt{T}\,\widehat{\boldsymbol{\rho}}_T(l)\ \xrightarrow{d}\ \mathcal{N}_{P^2B^2}(\mathbf{0}_{P^2B^2},\Omega),
\end{equation*}
for some symmetric, positive semidefinite covariance matrix $\Omega$, being $\boldsymbol{0}_{P^2B^2}$ the $P^2B^2$-dimensional vector with all zeros. A natural estimator $\widehat{\Omega}$ can be obtained from the sample covariance of $\sqrt{T}\,\widehat{\boldsymbol{\rho}}_T(l)$, for instance, via plug-in estimators of the relevant second-order moments of the indicator functions. Let $\widehat{\lambda}_1,\ldots,\widehat{\lambda}_{P^2B^2}$ denote the eigenvalues of $\widehat{\Omega}$. Then, given the data, the random variable
\begin{equation*}
\sum_{j=1}^{P^2B^2}\widehat{\lambda}_j Y_j^2,\qquad Y_j\overset{\text{i.i.d.}}{\sim}\mathcal{N}(0,1),
\end{equation*}
provides a consistent approximation of the null distribution of $T\,\widehat{S}_T(l)$.

An asymptotically level-$\alpha$ omnibus test for
\begin{equation*}
H_{0,l}^{\mathrm{omni}}:\ \rho(\tau,\tau',l,\beta,\beta') = 0 \quad \text{for all } (\tau,\tau')\in\mathcal{T}^2,\ (\beta,\beta')\in\mathcal{B}^2,
\end{equation*}
rejects $H_{0,l}^{\mathrm{omni}}$ whenever $T\,\widehat{S}_T(l) > c_{1-\alpha}(l)$, where $c_{1-\alpha}(l)$ is the $(1-\alpha)$ empirical quantile of $\sum_{j=1}^{P^2B^2}\widehat{\lambda}_j Y_j^2$. By Theorem~\ref{th5}, $\widehat{S}_T(l)\xrightarrow{p} S_T(l)$, so the proposed test is consistent against any fixed alternative with $S_T(l)>0$.

A portmanteau-type extension can also be considered in the omnibus setting by aggregating the corresponding statistics over a collection of lags $l=1,\ldots, L$, in direct analogy with the fixed-level and threshold case discussed above. 

In the analyses carried out in the rest of the paper, we focus on the omnibus test statistic $\widehat{S}_T(l)$ for a fixed lag $l$, since, in practice, interest lies in assessing serial dependence at a given lag by aggregating information across a range of quantile levels and thresholds, rather than testing at a single combination $(\tau,\tau',\beta,\beta')$ or over multiple lags simultaneously.

\section{Simulation study}\label{sectionsimulations}

We examine the finite-sample performance of the proposed FQA-based omnibus test through a comprehensive simulation study. We begin by outlining the test implementation details, then introduce several benchmark procedures from the functional time series literature for comparison. The subsequent analyses investigate the empirical size of the proposed test under various i.i.d.\ functional settings, evaluate its power across different forms of serial dependence, and assess its computational efficiency.

\subsection{Implementation details for the omnibus test}\label{subsectionimplementation}

In all empirical analyses, we simplify the statistical construction by identifying quantile levels and thresholds. Specifically, we work with a single grid of probability levels $\mathcal{T}=\{\tau_1,\ldots,\tau_P\}\subset(0,1)$ and restrict attention to terms of the form $\widehat{\rho}(\tau_i,\tau_j,l,\tau_i,\tau_j)$, $1\le i,j\le P$, so that the thresholds $(\beta_k,\beta_\ell)$ coincide with the corresponding quantile levels $(\tau_i,\tau_j)$. In this way, the omnibus statistic can be written as $\widehat{S}_T(l)=\sum_{i,j=1}^P \widehat{\rho}^2(\tau_i,\tau_j,l,\tau_i,\tau_j)$, which reduces the dimensionality of the problem while still aggregating information across a rich collection of probability levels. Note that all methodological and asymptotic results derived for the general omnibus statistic continue to apply to this reduced case. Throughout, we take $\mathcal{T}=\{0.05,0.10,\ldots,0.90,0.95\}$, which covers central and moderately extreme quantiles without putting excessive weight on extremely small or large probability levels. Note that this grid is chosen in a generic manner, aiming to cover a broad range of quantile levels without adapting to the specific dependence structure of the time series under analysis. Consequently, the empirical power reported for this implementation can be interpreted as a conservative, worst-case benchmark, since a more tailored choice of quantile levels could only improve the detection of serial dependence. Ideally, one would design a data-driven selection procedure that adapts $\mathcal{T}$ to the underlying process in order to maximize power for each application, but developing such an adaptive scheme lies beyond the scope of this paper and is left for future research.

Let us explain the motivation behind our choice of the reduced test statistic as described above. Note that, for a fixed $\tau \in (0,1)$, the functional process $\{\mathcal{X}_t,\, t \in \mathbb{Z}\}$ induces two processes
\begin{equation*}
\bigl\{\mathcal{A}_\tau^t : t\in\mathbb{Z}\bigr\}
\quad\text{and}\quad
\bigl\{\mathcal{L}(\mathcal{A}_\tau^t) : t\in\mathbb{Z}\bigr\},
\end{equation*}
which are set-valued and real-valued, respectively. For a fixed $\beta \in(0,1)$, the probability $\mathbb{P}\big(\mathcal{L}(\mathcal{A}_\tau^t) \le \beta\big)$ is simply the distribution function of the stationary process of Lebesgue measures $\{\mathcal{L}(\mathcal{A}_\tau^t) : t\in\mathbb{Z}\}$ evaluated at $\beta$. For illustration, let us examine a simple example with $\tau=0.99$ and $\beta=0.01$. In this context, consider a realization of the original functional process together with the corresponding realization of $\{\mathcal{L}(\mathcal{A}_\tau^t) : t\in\mathbb{Z}\}$. Since we are combining a very high quantile level with a very small threshold, it is quite likely that none of the observed Lebesgue-measure values falls below $\beta$, so the empirical estimator of $\mathbb{P}\big(\mathcal{L}(\mathcal{A}_\tau^t) \le \beta\big)$ becomes zero, which in turn causes numerical issues when computing $\widehat{\rho}(\tau, \tau', l, \beta, \beta')$ (see Section~\ref{sectionbackground}). Although this is an extreme example, in practice, similar problems can arise for less pronounced discrepancies between quantile levels and thresholds. A rigorous study of the distribution of $\{\mathcal{L}(\mathcal{A}_\tau^t) : t\in\mathbb{Z}\}$ under specific functional time series models could clarify these phenomena further, but such analyses are beyond the scope of this paper and are left for future work.

Given a fixed lag $l$, the omnibus statistic is computed considering the vector $\widehat{\boldsymbol{\rho}}_T=\widehat{\boldsymbol{\rho}}_T(l)\in\mathbb{R}^{P^2B^2}$, obtained by stacking the corresponding FQA estimators in a fixed order (see the proof of Theorem~\ref{th4}), and the test statistic $T\,\widehat{S}_T(l)$ is used as the basis for inference. To approximate the null distribution of $T\,\widehat{S}_T(l)$ by a weighted sum of independent $\chi^2$ variables (refer to Theorem~\ref{th4}), we estimate the covariance matrix $\Omega$ in the Gaussian limit of $\sqrt{T}\,\widehat{\boldsymbol{\rho}}_T$ consistently. Concretely, we compute a consistent estimator $\widehat{\Omega}$ based on the underlying indicator-based quantities over time, using a lag window mechanism to stabilize the contribution of nonzero autocovariances.

Let $\widehat{\lambda}_1,\ldots,\widehat{\lambda}_{P^2B^2}$ denote the resulting eigenvalues of $\widehat{\Omega}$. We approximate critical values for $T\,\widehat{S}_T(l)$ by Monte Carlo simulation. Specifically, we generate i.i.d.\ standard normal variables $Y_{j,m}\sim\mathcal{N}(0,1)$ and compute
\[
Q_m=\sum_{j=1}^{P^2B^2}\widehat{\lambda}_j Y_{j,m}^2,\qquad m=1,\ldots,M,
\]
being $M$ the number of Monte Carlo replicates (taken as $M=10000$ in our implementation), and use the empirical $(1-\alpha)$ quantile of $\{Q_m\}_{m=1}^M$ as $c_{1-\alpha}(l)$.

\subsection{Competing tests}\label{subsectioncompeting}

For comparison, we consider four well-established approaches as competing methods: the test of independence proposed by \citet{gabrys2007portmanteau}, denoted FPC; a test based on the FACF \citep{horvath2013test,kokoszka2017inference}; the spectral-domain test of \citet{characiejus2020general}, denoted SDO; and the FSACF-based test of \citet{yeh2023functional}. A description of these tests, together with the implementation used in this paper, is provided in Section~\ref{sec:supp_2} of the Supplement.

\subsection{Finite-sample size under the null hypothesis}\label{subsectionsize}

To examine the finite-sample size of the proposed omnibus FQA test, we consider four distinct scenarios under the null hypothesis of strong white noise. Each scenario is characterized by a different data-generating mechanism for the functional time series, chosen to represent a range of marginal distributions and smoothness properties while preserving temporal independence. A description of the four scenarios is provided below:

\begin{asparaitem}
\item \textbf{Scenario 1.} We consider $\mathcal{X}_t(u) = \mathcal{W}_t(u)$, where $\{\mathcal{W}_t(u),\, t\in\mathbb{Z},\, u\in\mathcal{I}\}$ is a sequence of i.i.d.\ Brownian motions on $\mathcal{I}$.
\item \textbf{Scenario 2.} We consider i.i.d.\ Gaussian functional white noise. Specifically, for each $t \in \mathbb{Z}$ and $u\in\mathcal{I}$, the generating process is defined by $\mathcal{X}_t(u) = \varepsilon_t(u)$, where $\{\varepsilon_t(u),\, t\in\mathbb{Z},\, u\in\mathcal{I}\}$ are independent standard Gaussian random variables.
\item \textbf{Scenario 3.} We consider i.i.d.\ functional observations with a common quadratic mean function and heavy-tailed white noise. Specifically, for each $t \in \mathbb{Z}$ and $u \in \mathcal{I}$, the generating process is defined by $\mathcal{X}_t(u) = u^2 + \varepsilon_t(u)$, where $\{\varepsilon_t(u),\, t\in\mathbb{Z},\, u\in\mathcal{I}\}$ are independent Student-$t$ random variables with $3$ degrees of freedom.
\item \textbf{Scenario 4.} We consider i.i.d.\ functional observations generated from a Fourier--Cauchy process. For each $t \in \mathbb{Z}$ and $u \in \mathcal{I}$, the generating process is defined by
\begin{equation*}
\mathcal{X}_t(u) = Z_{1,t} + \sum_{k=1}^{3} \bigl\{ Z_{2k,t}\cos(2\pi k u) + Z_{2k+1,t}\sin(2\pi k u) \bigr\},
\end{equation*}
where $Z_{j,t}$ are i.i.d. standard Cauchy random variables. This scenario was used in the simulation study of \cite{yeh2023functional}.
\end{asparaitem}

These scenarios provide a diverse set of strong white noise models, ranging from continuous Brownian motion (Scenario~1) to discontinuous i.i.d.\ Gaussian and heavy-tailed curves (Scenarios~2 and 3), including Cauchy-driven processes (Scenario~4). They allow us to assess the proposed test under light- and heavy-tailed marginals, including settings where higher-order moments are poorly behaved or nonexistent, while preserving temporal independence. Note that scenarios with discontinuous paths pose particular challenges due to their roughness.

For a given scenario, we independently simulate $N=2000$ functional time series of length $T$ from the corresponding generating mechanism. We consider $p=500$ evenly spaced points in $[0,1]$ to create the observations for each curve. The proposed omnibus test and the four competing procedures are then applied to each series to test for serial dependence at lag $l=1$. Table~\ref{tablenull} reports the empirical rejection rates for each scenario and method for $T=100, 200, 500, 1000$ and significance levels $\alpha = 0.05$ and $\alpha = 0.01$.

Table~\ref{tablenull} shows that the proposed omnibus test achieves empirical sizes that are very close to the nominal levels across all scenarios and values of \(T\), including the shortest series length \(T = 100\), indicating reliable finite-sample performance under strong white noise. The FPC test displays a reasonable size in some configurations but tends to either overreject or underreject in others. An important limitation of this procedure is the need to predefine the number of principal components to retain, as different choices can lead to markedly different rejection rates. In contrast, the remaining three methods (FACF, SDO, and FSACF) display reasonable rejection rates in Scenario~1 but exhibit poor performance in scenarios with discontinuous sample paths (Scenarios~2 and~3). In Scenario~4, SDO systematically overrejects across all settings; FACF performs better than in Scenarios~2 and~3, though still far from optimal; and FSACF achieves rejection rates fairly close to the nominal levels for the largest series lengths ($T = 500, 1000$).
\begin{table}[!htb]
\centering
\caption{\small Empirical rejection rates of five tests (lag $l=1$) for four scenarios under the null hypothesis, for different values of the series length $T$ and nominal levels $\alpha=0.05, 0.01$.}\label{tablenull}
\begin{footnotesize}
\tabcolsep 0.037in
\renewcommand{\arraystretch}{1}
\begin{tabular}{@{}lccccccccccccccccc@{}}
\toprule
& \multicolumn{8}{c}{Scenario 1} & & \multicolumn{8}{c}{Scenario 2} \\
\cmidrule(lr){2-9}\cmidrule(lr){11-18}
 & \multicolumn{2}{c}{$T = 100$} & \multicolumn{2}{c}{200} & \multicolumn{2}{c}{500} & \multicolumn{2}{c}{1000}
    & & \multicolumn{2}{c}{$T = 100$} & \multicolumn{2}{c}{200} & \multicolumn{2}{c}{500} & \multicolumn{2}{c}{1000} \\
\cmidrule(lr){2-3}\cmidrule(lr){4-5}\cmidrule(lr){6-7}\cmidrule(lr){8-9}
\cmidrule(lr){11-12}\cmidrule(lr){13-14}\cmidrule(lr){15-16}\cmidrule(lr){17-18}
Test & $\alpha=0.05$ & 0.01 & 0.05 & 0.01 & 0.05 & 0.01 & 0.05 & 0.01
         & & 0.05 & 0.01 & 0.05 & 0.01 & 0.05 & 0.01 & 0.05 & 0.01 \\
\midrule
FQA &   0.047   &   0.014   &  0.052 & 0.010     &  0.050    &  0.013   &  0.050    &   0.011  
         & & 0.054   &   0.013   &  0.051 & 0.010     &   0.045   &   0.010  &  0.055    &   0.009   \\
FPC      &   0.044   &  0.006    &  0.045    &   0.004   &    0.050  &   0.011   &  0.043    & 0.011      
         &  &  0.040   &  0.007    &  0.042    &   0.007   &    0.045  & 0.007     &  0.055    &   0.011    \\
FACF     &   0.040   &   0.009   & 0.052     &   0.011   &   0.047   &   0.012   &  0.048    & 0.016      
         &  & 0.000   &   0.000   & 0.000     &   0.000   &  0.002    & 0.000     &   0.006   &   0.000    \\
SDO      &   0.048   &   0.011   &  0.054    &   0.015   &  0.059    &   0.014   &   0.055  & 0.014      
         &  &   0.402   &   0.005   &  0.401    &   0.022   &   0.266   &   0.029   &  0.166   &   0.025     \\
FSACF    &  0.040    &   0.006   &   0.047   & 0.008 &    0.051  &    0.011   &  0.050    &  0.010    
         & & 0.000    &   0.000   &   0.000   & 0.000 &  0.004    &   0.000    &   0.026   &     0.002    \\
\midrule
& \multicolumn{8}{c}{Scenario 3} & & \multicolumn{8}{c}{Scenario 4} \\
\cmidrule(lr){2-9}\cmidrule(lr){11-18}
 & \multicolumn{2}{c}{$T = 100$} & \multicolumn{2}{c}{200} & \multicolumn{2}{c}{500} & \multicolumn{2}{c}{1000}
    & & \multicolumn{2}{c}{$T = 100$} & \multicolumn{2}{c}{200} & \multicolumn{2}{c}{500} & \multicolumn{2}{c}{1000} \\
\cmidrule(lr){2-3}\cmidrule(lr){4-5}\cmidrule(lr){6-7}\cmidrule(lr){8-9}
\cmidrule(lr){11-12}\cmidrule(lr){13-14}\cmidrule(lr){15-16}\cmidrule(lr){17-18}
Test & $\alpha=0.05$ & 0.01 & 0.05 & 0.01 & 0.05 & 0.01 & 0.05 & 0.01
         & & 0.05 & 0.01 & 0.05 & 0.01 & 0.05 & 0.01 & 0.05 & 0.01 \\
\midrule
FQA &   0.049   &   0.011   &   0.055   &  0.010    &   0.053   &  0.010    &  0.047    &   0.010   
         & & 0.051   & 0.012     &   0.049   &  0.012    &  0.055    &   0.010   &   0.052   &  0.009    \\
FPC      &   0.076   &  0.062    &  0.063    &  0.047    &  0.034    &   0.030   &  0.040    &     0.032 
         & &  0.105  &   0.087   &   0.083   & 0.071     &   0.052   &   0.044   &   0.037   & 0.034     \\
FACF     &   0.006   &    0.004  & 0.050     &   0.027   &   0.092   &   0.048   &   0.110   &  0.074    
         & &  0.060  &   0.047   & 0.046     & 0.034     &   0.043   &   0.035   &  0.034    & 0.028      \\
SDO      &  0.344    &  0.162    &  0.372    &   0.209   &  0.354    &   0.228   &   0.296   &  0.182    
         & &  0.092  &   0.071   &   0.088   & 0.074     &   0.094   &   0.086   &   0.098   & 0.087     \\
FSACF    &   1.000   &  1.000    &   1.000   & 1.000      &  1.000    &   1.000   &   1.000   &  1.000    
         & & 0.040   &  0.006    & 0.050 & 0.001 & 0.061     &   0.014   &   0.066   &   0.014   \\
\bottomrule
\end{tabular}
\end{footnotesize}
\end{table}

\subsection{Finite-sample power under serial dependence}\label{subsectionpower}

To assess the finite-sample power of the proposed test, we study two additional scenarios under a variety of serial dependence structures and error distributions. Each scenario is defined by a different data-generating mechanism for the functional time series, designed to capture diverse patterns of temporal dependence and marginal behavior. The mechanisms are based on the functional autoregressive (FAR) model of order \(p\), which we introduce now. Let \(\{\mathcal{X}_t, t \in \mathbb{Z}\}\) be a sequence of random functions in \(L^2(\mathcal{I})\). We say that \(\mathcal{X}_t\) follows a FAR model of order \(p\), FAR\((p)\), if
\begin{equation*}
\mathcal{X}_t(u)
\;=\;
G\bigl(\mathcal{X}_{t-1},\dots,\mathcal{X}_{t-p}\bigr)(u)
\;+\;
\varepsilon_t(u),
\qquad u \in \mathcal{I},
\end{equation*}
where $G(\cdot)$ is an operator from \((L^2(\mathcal{I}))^p\) into \(L^2(\mathcal{I})\), and \(\{\varepsilon_t, t \in \mathbb{Z}\}\) is an i.i.d.\ sequence of \(L^2(\mathcal{I})\)-valued innovations with mean zero. A description of the particular form of FAR$(p)$ considered in each one of the scenarios is provided below:

\begin{asparaitem}
\item \textbf{Scenario~5.} We consider a linear FAR(1) model, that is, we take
\(p=1\) and set \(G(\mathcal{X}_{t-1})(u) = \int_{0}^{1} \phi(u,v)\,
\mathcal{X}_{t-1}(v)\,dv\), so that \(\mathcal{X}_t(u) =
\int_{0}^{1} \phi(u,v)\,\mathcal{X}_{t-1}(v)\,dv + \varepsilon_t(u)\) for
\(u \in \mathcal{I}\). The kernel \(\phi\) is square–integrable, inducing a
Hilbert–Schmidt operator on \(L^2(\mathcal{I})\). We will refer to this scenario as FAR(1). 
\item \textbf{Scenario~6.} We consider a threshold FAR(1) model \citep{li2024functional}. Again, we take \(p=1\) and let the operator \(G(\cdot)\) switch between two linear integral
operators according to a threshold functional of the previous curve.
More precisely,
\[
\mathcal{X}_t(u)
=
\begin{cases}
\displaystyle \int_{0}^{1} \phi_1(u,v)\,\mathcal{X}_{t-1}(v)\,dv
+\varepsilon_t(u),
& \text{if } r(\mathcal{X}_{t-1}) \le s,\\[6pt]
\displaystyle \int_{0}^{1} \phi_2(u,v)\,\mathcal{X}_{t-1}(v)\,dv
+\varepsilon_t(u),
& \text{if } r(\mathcal{X}_{t-1}) > s,
\end{cases}
\qquad u \in \mathcal{I},
\]
where \(\phi_1(\cdot,\cdot)\) and \(\phi_2(\cdot,\cdot)\) are square–integrable kernels, \(r(\cdot)\) is a real–valued threshold functional on
\(L^2(\mathcal{I})\), and \(s\) is a fixed threshold constant.
This specification induces regime–switching nonlinear dependence while
remaining within the FAR(1) framework. We will refer to this scenario as TFAR(1).
\end{asparaitem}

The power analyses are carried out as follows. Both scenarios employ the common Gaussian kernel $\phi^*(u,v)=c\exp(-(u^2+v^2)/2)$, that is, $\phi=\phi_1=\phi_2=\phi^*$. This kernel choice is consistent with the related literature; see, e.g., \citet{kokoszka2017inference,yeh2023functional,lopezoriona2025}. We fix the series length at $T=200$ and consider $p=500$ evenly spaced points in $[0,1]$ to create the observations for each curve. For each scenario, we study the power of the different methods at this fixed length across a range of parameter values that determine the strength of the serial dependence. All tests are carried out considering the lag $l=1$. In each setting, power is computed as the proportion of rejections among 2000 independent realizations at a significance level $\alpha=0.05$. Four error distributions are analyzed: 
\begin{inparaenum}
\item[(i)] $\{\varepsilon_t(u),\, t\in\mathbb{Z},\, u\in\mathcal{I}\}$ are independent standard Gaussian random variables, 
\item[(ii)] $\{\varepsilon_t(u),\, t\in\mathbb{Z},\, u\in\mathcal{I}\}$ are independent Student-$t$ random variables with 3 degrees of freedom, 
\item[(iii)] $\{\varepsilon_t, t \in \mathbb{Z}\}$ is an i.i.d.\ sequence of Brownian motions, and 
\item[(iv)] the same as (iii) but with 10\% of randomly selected curves contaminated at 10\% of randomly selected grid points by adding spikes of height~10.
\end{inparaenum}

\begin{figure}[!htb]
\centering
\includegraphics[width=\textwidth]{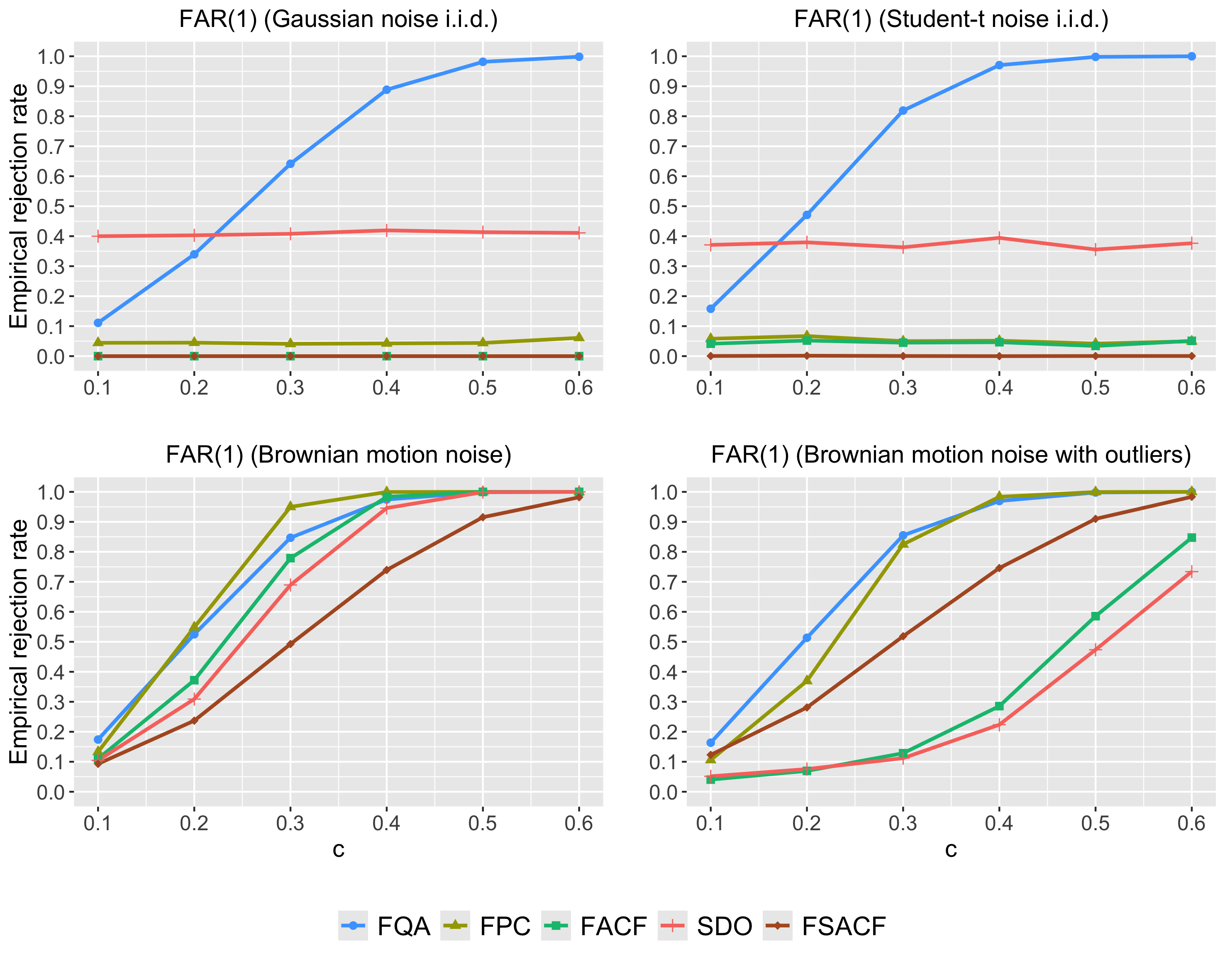}
\caption{\small Empirical rejection rates of five tests (lag $l=1$) for four settings under the alternative hypothesis (FAR(1) scenario), considering different noise distributions. Rejection rates are plotted as a function of a parameter ($c$) controlling the strength of serial dependence. A significance level of $\alpha=0.05$ is considered.}\label{power1}
\end{figure}

Empirical rejection rates for the FAR(1) scenario are shown in Figure~\ref{power1}, where each panel corresponds to a particular error distribution. Rejection rates are plotted as a function of the parameter $c$ in the kernel $\phi^*$. For Gaussian and Student-$t$ noise (top panels), our omnibus test attains perfect rejection rates when $c=0.6$, while most alternative tests fail to detect serial dependence for all values of $c$. The spectral test SDO shows rejection rates of around 0.40 across all values of $c$, but similar rejection rates were observed under the null hypothesis with Gaussian and Student-$t$ innovations for $T=200$, $\alpha=0.05$ (see Table~\ref{tablenull}).

For Brownian motion noise (bottom-left panel), all power curves exhibit reasonable behavior, with empirical power tending to 1 as $c$ increases across all tests. Among the five tests, FPC shows the highest rejection rates, followed closely by the FQA approach. When the time series is contaminated with outliers (bottom-right panel), the FQA test demonstrates robustness, with rejection rates very similar to those in the noncontaminated case. A similar pattern holds for FSACF, which is expected given its use of a robust autocorrelation measure. In contrast, the power of the remaining three tests (FPC, FACF, and SDO) declines substantially when outliers are introduced, with FACF and SDO showing particularly sharp drops in power. 

\begin{figure}[!htb]
\centering
\includegraphics[width=0.97\textwidth]{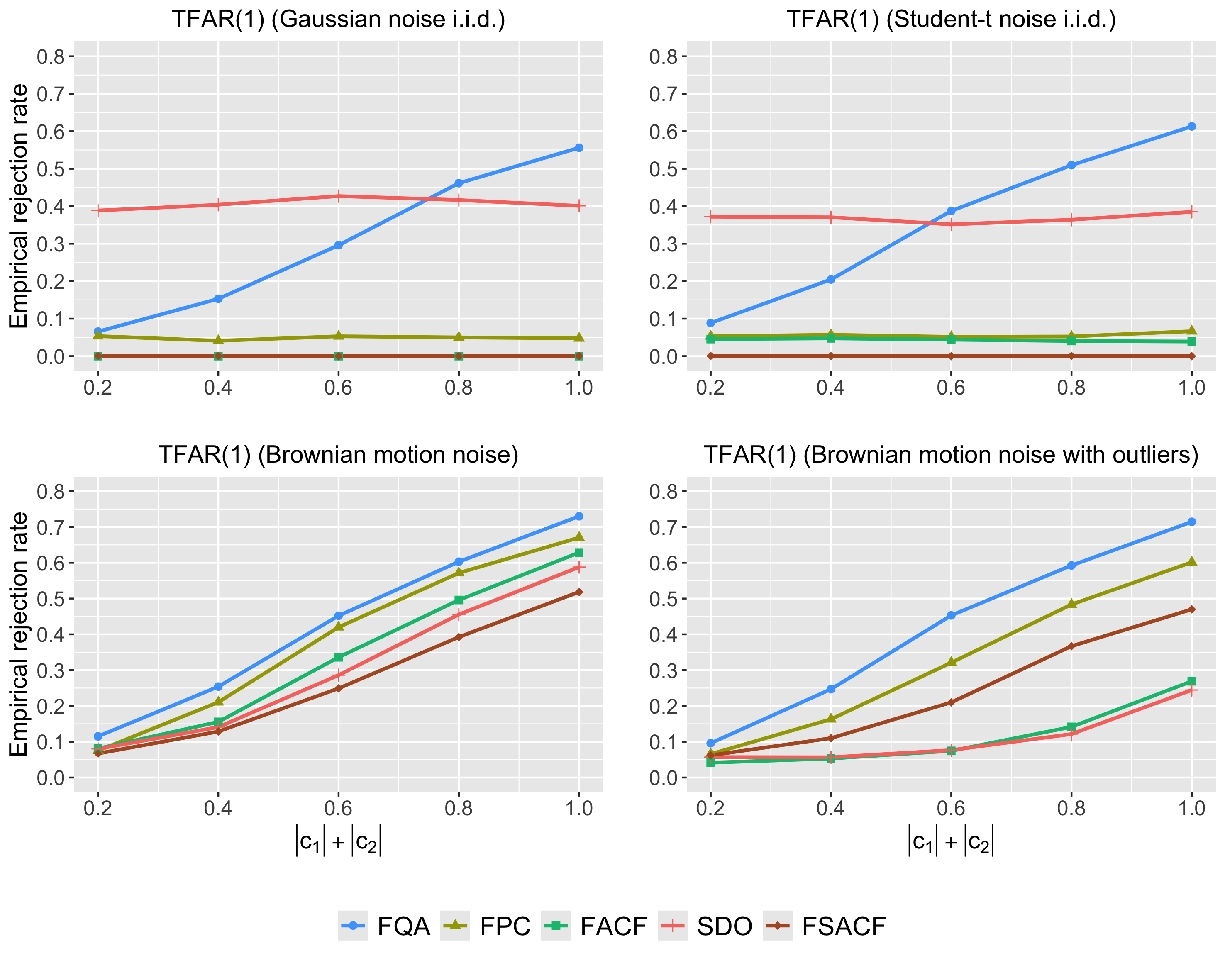}
\caption{\small Empirical rejection rates of five tests (lag $l=1$) for four settings under the alternative hypothesis (TFAR(1) scenario), considering different noise distributions. Rejection rates are plotted as a function of a quantity ($|c_1| + |c_2|$) controlling the strength of serial dependence. A significance level of $\alpha=0.05$ is considered.}\label{power2}
\end{figure}

Empirical rejection rates for the TFAR(1) scenario are shown in Figure~\ref{power2}. Rejection rates are plotted as a function of $|c_1| + |c_2|$, where $c_1$ and $c_2$ denote the constants associated with the kernels $\phi_1$ and $\phi_2$, respectively. The underlying simulation mechanism consists of the following steps: 
\begin{inparaenum}
\item[(i)] setting a positive value $C$ such that $|c_1| + |c_2| = C$, 
\item[(ii)] at each simulation replicate, generating $c_1$ uniformly distributed between $0$ and $C$, and 
\item[(iii)] setting $c_2 = c_1-C$. 
\end{inparaenum}
We consider values of~$C$ between $0$ and $1$, which guarantees that all processes used in this scenario are strictly stationary, since $\max\big\{\|\phi_1\|_\mathcal{H}, \|\phi_2\|_\mathcal{H}\big\}<1$; see Theorem~1 and Corollary~1 in \cite{li2024functional}. In all cases, we set the threshold $s = 0$ and the functional $r(f) = \int_0^1 f(u) \, du$ for $f \in \mathcal{H}$.

Figure~\ref{power2} shows that the proposed omnibus test outperforms all alternatives across all settings. The SDO test once again exhibits rejection rates that are largely insensitive to the strength of serial dependence under i.i.d.\ Gaussian or Student-\(t\) noise, whereas the remaining alternatives display very poor empirical power in these cases. Similar to the FAR(1) scenario, all competing tests improve their rejection rates when i.i.d.\ noise is replaced by Brownian motion. The proposed test again demonstrates robustness to outliers, whereas the rejection rates of FACF and SDO clearly deteriorate in this situation.

\begin{figure}[!htb]
\centering
\includegraphics[width=0.96\textwidth]{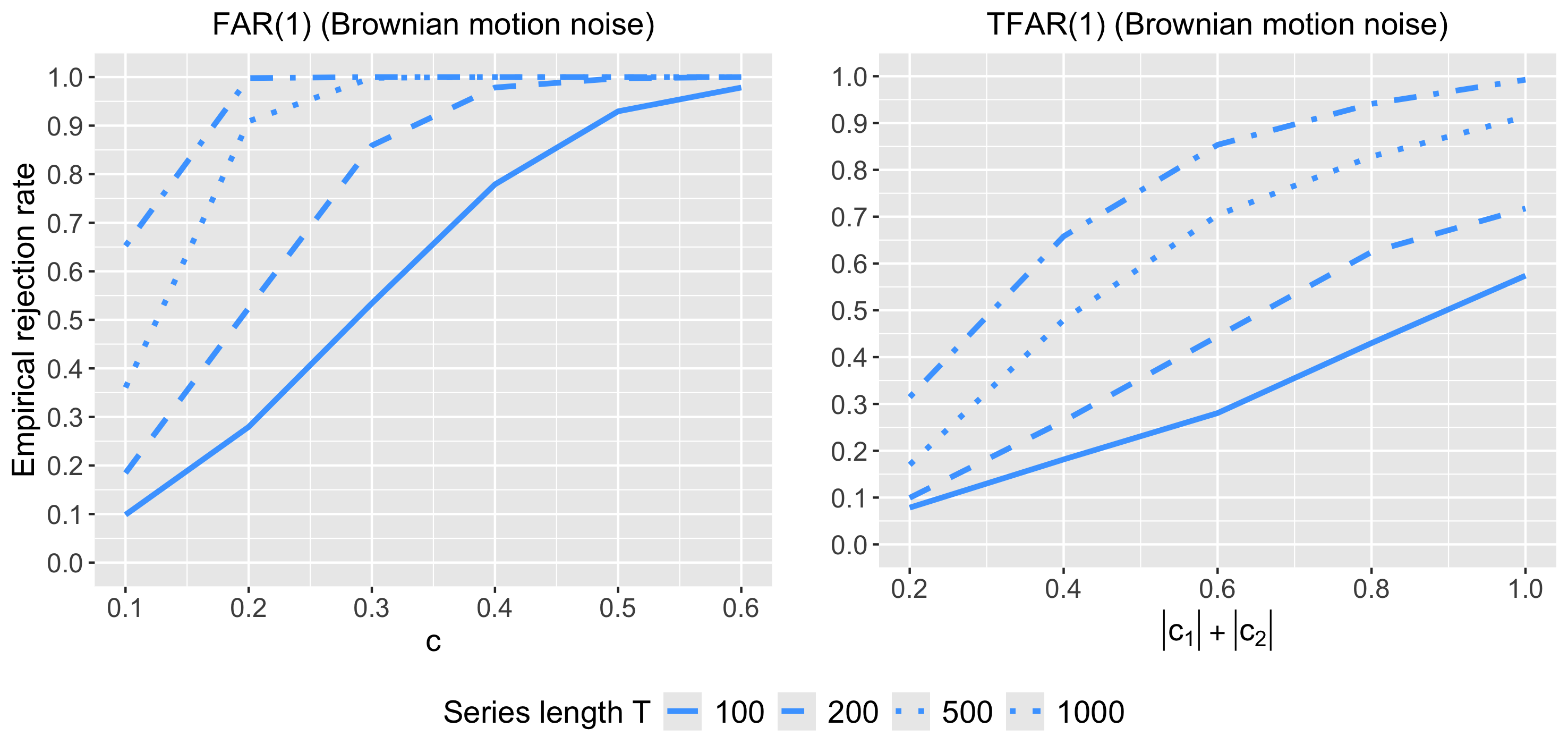}
\caption{\small Empirical rejection rates of the proposed omnibus test (lag $l=1$) for two settings under the alternative hypothesis (FAR(1) and TFAR(1) scenarios with Brownian motion noise), for different values of $T$. Rejection rates are plotted as a function of the parameters controlling the strength of serial dependence. A significance level of $\alpha=0.05$ is considered.}\label{powert}
\end{figure}

To examine the empirical power of the proposed omnibus test with respect to the series length~$T$, we replicate the above experiments for test FQA, but considering different values of $T$ in addition to $T=200$. All remaining elements are kept the same. For simplicity, for both scenarios, we focus on the results for the case in which the noise process $\{\varepsilon_t, t \in \mathbb{Z}\}$ is an i.i.d.\ sequence of Brownian motions, but analogous conclusions can be reached in the other settings. Empirical rejection rates for the FAR(1) and TFAR(1) scenarios as a function of the parameters controlling the degree of serial dependence are given in the left and right panels of Figure~\ref{powert}, respectively, for different values of $T$. We observe that larger values of $T$ lead to higher rejection rates. In all settings, rejection rates tend to one as $T$ increases. This analysis empirically confirms the consistency of the proposed omnibus statistic stated in Theorem~\ref{th5}.

In summary, the simulation experiments demonstrate that the proposed omnibus test constitutes a powerful and robust tool for strong white noise testing in functional time series. Notably, the test exhibits reliable behavior and substantial power even in relatively simple scenarios where several classical procedures perform poorly, such as when the data-generating processes produce rough, noncontinuous sample paths driven by i.i.d. Gaussian or Student-\(t\) noise. Although these designs depart from the smooth-curve idealization commonly assumed in functional data analysis, they are natural in practice when functions are densely observed but contaminated by pointwise measurement error or heavy-tailed perturbations, as in high-frequency financial data or noisy spectrometric and biomedical signals. These challenging settings also serve as stringent benchmarks for evaluating robustness in the presence of roughness and outliers. In such contexts, the proposed omnibus statistic, constructed from quantile-based excursion sets rather than relying solely on second-order structure, achieves accurate size and high power, effectively mitigating the deterioration observed for some existing tests. The consistent performance across settings with varying noise distributions, path properties, and dependence structures illustrates that the methodology is genuinely robust and broadly applicable, without being constrained to idealized smooth functional structures. This versatility underscores the practical relevance of the FQA-based approach and its suitability for real-world functional time series analysis.

\subsection{Computational time}\label{subsectionct}

To empirically examine the running time of the proposed omnibus test and the alternative tests, we revisit Scenario~2 (i.i.d.\ Gaussian white noise) described above, chosen for its simplicity since running times are similar across all considered scenarios. Specifically, we consider the corresponding simulation experiment outlined in Section~\ref{subsectionsize}. For each value of the series length~$T$, we record the average runtime required by the proposed approach and the alternative techniques to compute the corresponding test statistic and critical value. All computations are performed on a MacBook Pro equipped with an Apple M2 Pro chip and 16~GB of RAM, and all programs are implemented and executed in the \proglang{R} software \citep[][version~4.4.3]{rjournal}.

The results are shown in Table~\ref{tabletime}. The proposed omnibus test exhibits reasonable computational times across all values of $T$, with a much less pronounced increase in runtime with $T$ than some alternative approaches. For instance, for a functional time series of length $T=1000$, the proposed test runs in approximately two seconds. In contrast, the FSACF shows by far the highest runtimes among all methods across the considered values of $T$.

\begin{table}[!htb]
\centering 
\caption{\small Average runtime (in seconds) of different approaches for computing the test statistic and the corresponding critical value in Scenario~2 for various series lengths $T$.}\label{tabletime}
\tabcolsep 0.4in
\renewcommand{\arraystretch}{1}
\begin{small}
\begin{tabular}{@{}lrrrrr@{}}
\toprule
$T$ & FQA & FPC & FACF & SDO & FSACF   \\ \midrule
100	&     0.774          &     0.045                 & 0.113  & 0.012 & 10.973   \\
200	&    0.921            &  0.142                  & 0.212 & 0.071 & 27.325  \\
500	&   1.468            &   0.734                 & 0.448  & 0.704  &  60.833 \\
1000	&   2.153               & 4.011                     & 0.833 &  5.011 &  115.006 \\ 
\bottomrule
\end{tabular}
\end{small}
\end{table}

\section{Application to intraday stock returns}\label{sectionrealdata}

We analyze 5-minute closing prices for six S\&P~500 companies—Alphabet Class~A (GOOGL), AT\&T~(T), Cisco Systems (CSCO), Comcast (CMCSA), Electronic Arts (EA), and Meta Platforms (META)--recorded between 9:30~a.m.\ and 4:00~p.m.\ (EST) from January~2,~2018, to December~31,~2020. The data were obtained from Refinitiv Datascope (\url{https://select.datascope.refinitiv.com/DataScope/}). For each company, the intraday prices are treated as daily price curves, yielding six functional time series over the 2018–2020 period. With 251, 252, and 253 trading days in 2018, 2019, and 2020, respectively, each series has length $T=756$, and each curve is observed at $p=78$ equally spaced time points corresponding to 5-minute intervals during the 6.5-hour trading day.

Since the functional time series of prices are nonstationary, they are transformed into intraday log-returns. Let $P_{i,t}(u_j)$ be the intraday 5-minute closing price of the $i$\textsuperscript{th} company at time $u_j$ on trading day $t$, the corresponding sequence of log-returns is defined as: $R_{i,t}(u_j)=\ln P_{i,t}(u_j)-\ln P_{i,t}(u_{j-1})$, where $i=1,\dots,6$, $j=2,\dots,78$ and $t=1,\dots,756$. Figure~\ref{plotstocks_supp} in Section \ref{sec:supp_3} of the Supplement displays the first 100 curves (corresponding to 2018) of the functional time series of log-returns for the six companies.

Although the values of the $x$-axis in these graphs were chosen according to the daily trading times, the functional time series of logarithmic returns is assumed to be defined at 77 evenly spaced points in the interval $[0,1]$. The plots reveal evidence of outliers and heavy-tailed behavior, with extreme log returns observed across all companies. These properties are typical of high-frequency financial data and motivate the proposed omnibus test, which is robust to outliers and the distributional form of errors. The six functional time series considered here were also analyzed by \citet{lopezoriona2025} for the fuzzy clustering of S\&P~500 companies based on the FQA features.

The proposed omnibus test is applied to each of the six functional time series to test for serial dependence at lags $l=1,\dots,10$. The implementation details follow those described in Section~\ref{subsectionimplementation}, with $p$-values approximated by Monte Carlo simulation using $M=10000$ replicates of the linear combination of $\chi^2$ variables. Such dependence diagnostics are standard in functional time series analysis, typically performed prior to model fitting to assess the extent of temporal structure, guide model selection, or conduct exploratory analyses of persistence patterns across multiple lags. 

For comparison purposes, the alternative tests considered in the simulations are also applied to the financial time series. For the FPC, FACF, and FSACF tests, serial dependence is tested using a portmanteau test, since this is the implementation considered in the employed \proglang{R} functions (see Section \ref{sec:supp_2} in the Supplement). Specifically, for lag $l$, these tests assess the joint null hypothesis of no serial dependence across lags $1$ to $l$, against the alternative that dependence exists at some lag in this range. The SDO method, being a spectral domain-based method, does not depend on a specific lag. For the FPC, the number of principal components is set to three, consistent with the simulation~study.

The $p$-values produced by all the tests are reported in Table~\ref{tablepvalues}. The proposed omnibus test yields $p$-values numerically equal to zero in all settings, thus consistently detecting significant serial dependence at each lag from $l=1$ to $10$. The results of the competing procedures are more heterogeneous. For example, the FPC and FACF tests detect significant serial dependence at all considered lags for the functional time series of companies T and EA, but fail to do so at every lag for CSCO. The spectral test SDO identifies significant serial dependence (numerical $p$-value equal to zero) in three of the six functional time series, but not in the remaining three. The FSACF test rejects the null hypothesis of independence in most configurations, yet produces $p$-values above the conventional significance level $\alpha=0.05$ in several cases (for instance, at several lags for the CMCSA series).

\begin{footnotesize}
\begin{center}
\tabcolsep 0.13in
\renewcommand{\arraystretch}{1}
\begin{longtable}{@{}c *{5}{c}c *{5}{c}@{}}
\caption{\small The $p$-values of tests for the null hypothesis of strong white noise at lags $l \in \{1, 2, \ldots, 10\}$ for six functional time series of log-returns. $p$-values below 0.05 are highlighted in bold. The SDO test does not depend on the lag $l$, so only one $p$-value is reported for each time series.}
\label{tablepvalues} \\
\toprule
\endfirsthead
\endhead
\midrule
\multicolumn{12}{r}{Continued on next page} \\ 
\endfoot
\endlastfoot
\multicolumn{6}{c}{GOOGL} & \multicolumn{6}{c}{$T$} \\
\cmidrule(lr){2-6}\cmidrule(lr){7-12}
$l$ & FQA & FPC & FACF & SDO & FSACF & $l$ & FQA & FPC & FACF & SDO & FSACF \\
\hline
1  &  \textbf{0.000}  &  1.000   &   0.935  &  \textbf{0.000}    &  \textbf{0.001}   & 1  & \textbf{0.000} &  \textbf{0.000}   &  \textbf{0.000}   &  \textbf{0.000}    &  \textbf{0.000}   \\
2  & \textbf{0.000} &  1.000   &  1.000   &   -   &  \textbf{0.000}   & 2  & \textbf{0.000}     &   \textbf{0.000}  &  \textbf{0.000}   &  -    & \textbf{0.000}     \\ 
3  &  \textbf{0.000}   &  1.000   & 1.000    &  -    &  \textbf{0.000}   & 3  & \textbf{0.000}     &  \textbf{0.000}   &  \textbf{0.000}   &  -    & \textbf{0.000}    \\
4  &  \textbf{0.000}   &  1.000   &  \textbf{0.000}   &  -    &  \textbf{0.000}   & 4  & \textbf{0.000}     &  \textbf{0.000}   &  \textbf{0.000}   &  -    & \textbf{0.000}    \\
5  &  \textbf{0.000}   &   1.000  &  \textbf{0.000}   &  -    &  \textbf{0.000}   & 5  & \textbf{0.000}     &  \textbf{0.000}   & \textbf{0.000}    &  -    & \textbf{0.000}    \\
6  &  \textbf{0.000}   &  1.000   &  \textbf{0.000}   &  -    &  \textbf{0.000}   & 6  & \textbf{0.000}    &   \textbf{0.000}  &  \textbf{0.000}   &  -    &  \textbf{0.000}   \\
7  &  \textbf{0.000}   &  \textbf{0.000}   &  \textbf{0.000}   &  -    &  \textbf{0.000}   & 7  & \textbf{0.000}     &  \textbf{0.000}   &  \textbf{0.000}   &  -    &  \textbf{0.000}   \\
8  &  \textbf{0.000}   &  \textbf{0.000}   &  \textbf{0.000}   &  -    &  \textbf{0.000}   & 
8  & \textbf{0.000}     &  \textbf{0.000}   &  \textbf{0.000}   &  -    &  \textbf{0.000}   \\
9  &  \textbf{0.000}   &  \textbf{0.000}   &  \textbf{0.000}   &  -    &  \textbf{0.000}   & 9  & \textbf{0.000}     & \textbf{0.000}    &  \textbf{0.000}   &  -    &  \textbf{0.000}   \\
10 &  \textbf{0.000}   &  \textbf{0.000}   &  \textbf{0.000}   &  -    &  \textbf{0.000}   & 10 & \textbf{0.000}     & \textbf{0.000}    &  \textbf{0.000}   &  -    &  \textbf{0.000}   \\
\hline
\multicolumn{6}{c}{CSCO} & \multicolumn{6}{c}{CMCSA} \\
\cmidrule(lr){2-6}\cmidrule(lr){7-12}
$l$ & FQA & FPC & FACF & SDO & FSACF & $l$ & FQA & FPC & FACF & SDO & FSACF \\
\hline
1  &  \textbf{0.000}   &  1.000   &   0.998  &  1.000    & \textbf{0.008}    & 1  &  \textbf{0.000}   &  1.000   &  1.000   &   0.153   &  0.385   \\
2  &  \textbf{0.000}   &  1.000   &   0.976  &  -    &  \textbf{0.003}   & 2  &  \textbf{0.000}   &  1.000   &  1.000   &  -    &  \textbf{0.012}   \\
3  &  \textbf{0.000}   &  1.000   &  1.000   &  -    &  \textbf{0.000}   & 3  &  \textbf{0.000}   &  1.000   &  1.000   &  -    &  \textbf{0.030}   \\
4  &  \textbf{0.000}   &  1.000   &  1.000   &  -    &  \textbf{0.000}   & 4  &  \textbf{0.000}   &  1.000   &  1.000   &  -    &  \textbf{0.022}   \\
5  &  \textbf{0.000}   &  1.000   &  1.000   &  -    &  \textbf{0.000}   & 5  &  \textbf{0.000}   &  1.000   &  1.000   &  -    &  \textbf{0.032}   \\
6  &  \textbf{0.000}   &  1.000   &  1.000   &  -    &  \textbf{0.000}   & 6  &  \textbf{0.000}   &  1.000   &  1.000   &  -    &  0.050   \\
7  &  \textbf{0.000}   &  1.000   &  1.000   &  -    &  \textbf{0.000}   & 7  &  \textbf{0.000}   &  1.000   &  1.000   &  -    &  0.075   \\
8  &  \textbf{0.000}   &  1.000   &  1.000   &  -    &  \textbf{0.000}   & 8  &  \textbf{0.000}   &  1.000   &  1.000   &  -    &  0.074   \\
9  &  \textbf{0.000}   &  1.000   &  1.000   &  -    &  \textbf{0.000}   & 9  &  \textbf{0.000}   &  1.000   &  \textbf{0.000}   &  -    &  0.103   \\
10 &  \textbf{0.000}   &  1.000   &  1.000   &  -    &  \textbf{0.000}   & 10 &  \textbf{0.000}   &  1.000   &  \textbf{0.000}   &  -    &   \textbf{0.030}  \\
\hline
\multicolumn{6}{c}{EA} & \multicolumn{6}{c}{META} \\
\cmidrule(lr){2-6}\cmidrule(lr){7-12}
$l$ & FQA & FPC & FACF & SDO & FSACF & $l$ & FQA & FPC & FACF & SDO & FSACF \\
\hline
1  &  \textbf{0.000}  &  \textbf{0.000}   &  \textbf{0.000}   &   \textbf{0.000}   &  \textbf{0.022}   & 1  &  \textbf{0.000}   &  1.000   &  \textbf{0.003}   &  1.000    & 0.149    \\
2  &  \textbf{0.000}   & \textbf{0.000}    &  \textbf{0.000}   &  -    &  \textbf{0.006}   & 2  &  \textbf{0.000}   &  1.000   &  0.797   &   -   &  0.193   \\
3  &  \textbf{0.000}   & \textbf{0.000}    &  \textbf{0.000}   &  -    &  \textbf{0.016}   & 
3  &  \textbf{0.000}   &  1.000   &  1.000   &  -    &  \textbf{0.002}   \\
4  &  \textbf{0.000}   & \textbf{0.000}   &   \textbf{0.000}  &   -   &  \textbf{0.001}   & 4  &  \textbf{0.000}   &  1.000   &  1.000   &  -    &  \textbf{0.000}   \\
5  &  \textbf{0.000}   & \textbf{0.000}    &  \textbf{0.000}   &  -    &  \textbf{0.000}   & 5  &  \textbf{0.000}   &  1.000   &   1.000  &  -    &  \textbf{0.000}   \\
6  &  \textbf{0.000}   & \textbf{0.000}    &  \textbf{0.000}   &  -    &  \textbf{0.000}   & 6  &  \textbf{0.000}   &  1.000   &  1.000   &  -    &  \textbf{0.000}   \\
7  &  \textbf{0.000}   & \textbf{0.000}    &  \textbf{0.000}   &  -    &  \textbf{0.000}   & 7  &  \textbf{0.000}   &  1.000   &  1.000   &  -    &  \textbf{0.000}   \\
8  &  \textbf{0.000}   & \textbf{0.000}    &  \textbf{0.000}   &  -    &  \textbf{0.000}   & 8  &  \textbf{0.000}   &  1.000   &  1.000   &  -    &  \textbf{0.000}   \\
9  &  \textbf{0.000}   & \textbf{0.000}    &  \textbf{0.000}   &  -    &  \textbf{0.000}   & 9  &  \textbf{0.000}   &  1.000   &  1.000   &  -    &  \textbf{0.000}   \\
10 &  \textbf{0.000}   & \textbf{0.000}    &  \textbf{0.000}   &  -    &  \textbf{0.000}   & 10 &  \textbf{0.000}   &  1.000   &  1.000   &  -    &  \textbf{0.000}   \\
\bottomrule
\end{longtable}
\end{center}
\end{footnotesize}

\vspace{-.5in}

In the six functional time series under study, we also applied the proposed omnibus statistic to test for serial dependence at lags beyond \(l = 10\). However, these additional results are omitted for simplicity. In most series, the strength of dependence exhibits a noticeable drop around lag \(l = 50\), but for several companies the test continues to detect significant dependence at substantially larger lags. To assess whether these long-lag rejections reflect genuine temporal structure, we repeated the analysis after randomly permuting the order of the intraday return curves for each company, thereby preserving the marginal properties while destroying temporal dependence; in these permuted series, the omnibus test did not reject the null hypothesis of strong white noise at the examined lags. This pattern is particularly meaningful in the context of high-frequency financial data \citep[see, e.g.,][]{li2020long}, where slowly decaying dependence in volatility and tail behavior, as well as episodes of sustained market turbulence or calm, can induce weak but persistent nonlinear and tail dependence over long horizons.

These analyses reveal that the proposed test better detects serial dependence in functional time series compared to several alternative procedures. This pattern holds in the challenging context of financial functional time series, which often exhibit complex dependence structures, outliers, and heavy tails. Consistent with the simulation study results, these data characteristics may explain why some competing methods fail to detect dependence. Notably, the FSACF test, which is robust to outliers and the lower-order moments of the data-generating process, is the only alternative that identifies significant serial dependence in most settings. These findings highlight the practical utility of the proposed test for explaining the dependence structure of financial functional time series before subsequent analyses, such as model fitting and forecasting.

To complement the above analyses, Section~\ref{sec:supp_3} of the Supplement presents additional graphical results for the FACF and FSACF tests. For each functional time series and method, plots display the test statistics and their associated critical values. Values at lag $l$ correspond to testing serial dependence at that individual lag (rather than the portmanteau testing up to $l$, as in Table~\ref{tablepvalues}). These supplementary results corroborate the findings from the main analysis.

\section{Conclusions and discussion}\label{sectiondiscussion}

This paper introduces a new class of time domain tests for i.i.d.\ white noise in functional time series based on FQA, going beyond procedures that assess only second-order dependence. By working with quantile-based excursion sets rather than classical covariance operators or other second-order summaries, the proposed omnibus statistic captures complex forms of serial dependence, including those concentrated in specific regions of the functional distribution, while remaining applicable under minimal assumptions beyond strict stationarity and performing well in the presence of heavy tails and outliers. 

The asymptotic results for the sample FQA and the omnibus statistic, under known and unknown quantile curves, provide a rigorous foundation for hypothesis testing under i.i.d.\ functional white noise, and guarantee consistency against a broad class of stationary and ergodic alternatives.

The simulation study shows that the proposed test exhibits accurate empirical size across a range of i.i.d.\ functional scenarios and often outperforms established competitors (such as tests based on functional principal components, autocovariance-based procedures, spectral tests, and robust methods relying on spherical autocorrelation), especially when the underlying dependence is nonlinear or not well-summarized by second-order features, or when heavy tails or outliers are present. Moreover, unlike alternative procedures, our test exhibits reliable performance both in scenarios that depart from the smooth-curve idealization commonly assumed in functional data analysis (where traditional methods fail), and in settings where the data-generating processes produce continuous sample paths, such as under Brownian motion noise. At the same time, the test’s implementation yields competitive computation times, rendering the proposed testing framework suitable for large-scale applications.

An application involving functional time series of intraday stock log-returns demonstrates the methodology's effectiveness in high-frequency financial settings with prevalent outliers and heavy tails. Testing for strong white noise is particularly relevant here, as it provides a comprehensive diagnostic for both linear and nonlinear serial dependence, which classical second-order tests (e.g., based on autocovariance operators) may miss due to their focus on mean-covariance structure amid conditional heteroscedasticity and heavy-tailed noise. In particular, the omnibus test reveals substantively meaningful deviations from i.i.d.\ behavior across multiple lags, providing a valuable complement to existing functional time series tools for model selection, risk assessment, and forecasting in finance.

This work provides early yet strong evidence on the use of FQA for independence testing in functional time series, and it opens several avenues for future research. First, it would be of interest to study the behavior of the omnibus test when quantile levels and thresholds are allowed to differ, and to develop data-driven procedures for selecting these quantities. Further directions include analyzing portmanteau versions of the proposed tests, deriving the power properties of the omnibus statistic under local alternatives to better understand its ability to detect weak signals, and conducting an in-depth assessment of the feasibility of the essential assumptions in Theorems~\ref{th3} and~\ref{th4}, with the goal of extending the framework to milder regularity conditions.

\vspace{-0.4cm}

\section*{Acknowledgments}

Ángel López-Oriona and Ying Sun thank KAUST for its support. Han Lin Shang thanks the financial support from an Australian Research Council Future Fellowship (FT240100338).

\vspace{-0.4cm}

\section*{Supplementary Materials}

The supplement file includes proofs of theorems and lemma in~S1, competing tests in~S2, and additional data analyses for the financial application in~S3.

\vspace{-0.4cm}

\section*{Code availability statement}

The \proglang{R} code necessary to run the main analyses presented in this paper is available in GitHub at \url{https://github.com/anloor7/PostDoc/tree/main/r\_code/white\_noise\_testing\_fts}.

\vspace{-0.4cm}

\section*{Data availability statement}

Data used in this paper were obtained from Refinitiv DataScope, accessible at \url{https://select.datascope.refinitiv.com/DataScope/}. 

\vspace{-0.4cm}

\section*{Disclosure statement}

The authors declare no conflicts of interest relevant to the content of this article.

\newpage	
{\normalsize
\bibliography{test_fqa.bib}
}
	
\end{document}